\documentclass[traditabstract]{aa} 
\usepackage{graphicx}
\usepackage[varg]{txfonts}
\usepackage{natbib}
\usepackage{lscape}
\usepackage[normalem]{ulem}
\usepackage{subfig}
\usepackage{caption}
\bibpunct{(}{)}{;}{a}{}{,} 

\begin{document}

   \title{Pure-hydrogen 3D model atmospheres of cool white dwarfs}

   \author{P.-E. Tremblay\inst{1}
          \and
          H.-G. Ludwig\inst{1}
          \and
          M. Steffen\inst{2}
          \and
          B. Freytag\inst{3}
          }

   \institute{Zentrum f\"ur Astronomie der Universit\"at Heidelberg, Landessternwarte, 
            K\"onigstuhl 12, 69117 Heidelberg\\
             \email{ptremblay@lsw.uni-heidelberg.de,hludwig@lsw.uni-heidelberg.de}
         \and
            Leibniz-Institut f\"ur Astrophysik Potsdam, An der Sternwarte 16, D-14482 Potsdam, Germany\\
             \email{msteffen@aip.de}
         \and
           Centre de Recherche Astrophysique de Lyon, UMR 5574: CNRS, Universit\'e de Lyon,
           \'Ecole Normale Sup\'erieure de Lyon, 46 all\'ee d'Italie, F-69364 Lyon Cedex 07, France \\
              \email{Bernd.Freytag@ens-lyon.fr}
             }

   \date{Received ..; accepted ..}
 
  \abstract {A sequence of pure-hydrogen CO5BOLD 3D model atmospheres of DA
    white dwarfs is presented for a surface gravity of $\log g = 8$ and
    effective temperatures from 6000 to 13,000~K. We show that convective
    properties, such as flow velocities, characteristic granulation size and
    intensity contrast of the granulation patterns, change significantly over
    this range. We demonstrate that these 3D simulations are not sensitive to
    numerical parameters unlike the 1D structures that considerably depend on
    the mixing-length parameters. We conclude that 3D spectra can be used
    directly in the spectroscopic analyses of DA white dwarfs. We confirm the
    result of an earlier preliminary study that 3D model spectra provide a
    much better characterization of the mass distribution of white dwarfs and
    that shortcomings of the 1D mixing-length theory are responsible for the
    spurious high-$\log g$ determinations of cool white dwarfs. In particular,
    the 1D theory is unable to account for the cooling effect of the
    convective overshoot in the upper atmospheres.}{}{}{}{}


   \titlerunning{Pure-hydrogen 3D model atmospheres of cool white dwarfs}
   \authorrunning{Tremblay et al.}
   \maketitle

\section{Introduction}

The multidimensional radiation-hydrodynamics (RHD) treatment of convective
motions in hydrogen-atmosphere DA white dwarfs was pioneered by the Kiel group
some fifteen years ago \citep{ludwig94,steffen95,freytag96}. These studies
have revealed that the differences between spectra computed on the basis of 2D
RHD simulations and those computed from standard 1D hydrostatic models were
small. However, recent research developments make it advisable to pursue the
computation of RHD models. First of all, three-dimensional RHD computations
(hereafter 3D simulations) became accurate enough to be the reference in
quantitative spectroscopy, in particular for the abundance determinations in
the Sun \citep{asplund09,caffau11}. Secondly, the spectroscopic technique in
DA white dwarfs, which consists in comparing the observed line profiles of the
Balmer series with the predictions of detailed model atmospheres
\citep{weidemann80,bergeron92a}, became more widely used, with larger samples
\citep{SDSS} and higher signal-to-noise ratio (S/N) observations
\citep{gianninas11}. For high quality observations (S/N $>$ 50), it is often
quoted that the relative uncertainties in the atmospheric parameters, the
effective temperature ($T_{\rm eff}$) and surface gravity ($\log g$, in cgs
units), are of the order of 1$\%$ \citep{LBH05,koester09b}. As a consequence,
white dwarfs have been used as spectroscopic standards to achieve
high-precision calibrations at the 1\% level \citep[see, e.g.,][]{bohlin00}.

It has been shown long ago that the spectroscopically determined masses of
cool DA white dwarfs ($T_{\rm eff} < 13,000$~K) were as much as 20$\%$ higher
than those of hotter DA white dwarfs \citep{bergeron90,bergeron92a}. This
discrepancy is now observed in the spectroscopic mass distribution of all
surveys of DA white dwarfs \citep{TB11,gianninas11}. Multiple attempts were
made to explain this behaviour \citep{koester09a,TB10} but were
unsuccessful. However, it was suggested that because convection becomes
significant in the photosphere almost exactly as the so-called high-$\log g$
problem becomes apparent, the 1D mixing-length theory \citep[][hereafter
  MLT]{MLT} was likely the culprit. Therefore, \citet[][hereafter Paper
  I]{paper1} computed the first four non-grey 3D model atmospheres of DA white
dwarfs (12,800 $> T_{\rm eff}$ (K) $>$ 11,300 and $\log g = 8$) using the
CO$^{5}$BOLD RHD code \citep{freytag12}. The code is usually applied to stars
with deep convective envelopes where it is not possible for the simulation
domain to span the full convection zone. However, the convection zone in these
hot white dwarfs is in fact thinner than the simulation domain in the vertical
direction.

The aim of Paper I was to predict quantitatively how the atmospheric
parameters are changed by using 3D spectra instead of the standard 1D
spectra. As a consequence, the methodological approach was to do a
differential analysis, by comparing both the spectra computed from 3D
structures, and 1D hydrostatic (LHD) structures \citep{caffau07} relying the
same microphysics and opacity bins as the 3D simulations, with standard 1D
spectra of white dwarfs \citep{TB11}. It was found that the 3D $\log g$
corrections had the correct amplitude to solve the high-$\log g$ problem for
DA white dwarfs in the range of $T_{\rm eff}$ studied in the paper.  It was
concluded that a weakness in the 1D MLT theory was creating the high-$\log g$
problem.

Several aspects of the 3D calculations can be improved in order to
understand to which extend these simulations can solve the high-$\log g$ problem,
and ultimately, to use them in the spectroscopic analyses of white
dwarfs. First of all, it is desirable to compute cooler 3D simulations, since the
high-$\log g$ problem is most prominent around $T_{\rm eff} = 10,000$~K (see,
e.g., Fig. 1 of Paper I). Furthermore, it is necessary to have 3D simulations
containing realistic microphysical properties in order to compare them
directly to white dwarf observations. For instance, the four 3D simulations
computed in Paper I rely on an equation-of-state (EOS) and opacity table
(including an opacity binning procedure) that have different microphysics in
comparison to standard 1D model atmospheres.

We tackle these shortcomings in this work by producing an improved sequence
of 12 DA model atmospheres, covering the range $13,000 > T_{\rm eff}$ (K) $>
6000$ at $\log g = 8$. In Sect.~2, we discuss about the precision of our 3D
simulations. In Sect.~3, we present the properties of our sequence of 3D
computations. Model spectra are computed in Sect.~4 and improved 3D $\log g$
corrections are derived in relation to the high-$\log g$ problem.  The
conclusion follows in Sect.~5.

\section{Establishing absolute properties}

The main objective of our study is to produce 3D model atmospheres that
represent real stellar conditions as closely as possible with the aim of
performing direct comparisons of these models with observations. We must
beforehand understand the sources of the differences between these 3D
structures and standard 1D structures, both computed with established
well-known codes. In other words, we must carefully compare the codes
producing these model atmospheres. Several groups have in the past computed
grids of 1D model atmospheres for DA white dwarfs
\citep[e.g.,][]{bergeron92a,finley97,kowalski06}.  Recent theoretical
developments include the improved modeling of the Lyman quasi-molecular
satellites \citep{allard04}, the calculation of the Ly-$\alpha$ red wing
opacity due to H$_2$-H collisions \citep{kowalski06} and the publication of
new Stark broadening profiles \citep{TB09} including in a consistent way the
non-ideal gas effects of \citet{hm88}. All of these improvements are now
included in our 1D model atmosphere code described in \citet{TB11} and
references therein. We note that the D.~Koester model atmosphere code includes
very similar physics at the time of publication \citep[see,
  e.g.,][]{girven11}. Hence, while we specifically use the \citet{TB11} grid
in this work, it is appropriate to describe them from now on as standard 1D
models.

\begin{figure}[!h]
\begin{center}
\includegraphics[bb=18 140 662 712,width=4in]{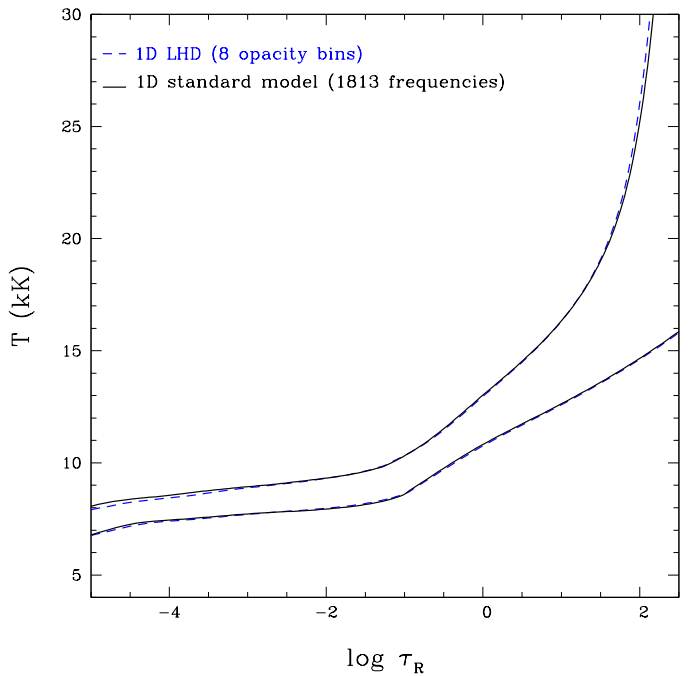}
\caption{Atmospheric temperature as a function of the logarithm of the
  Rosseland optical depth for DA models with effective temperatures of 8000
  (bottom) and 12,000~K (top). We show the standard 1D structures (solid black
  line) from \citet{TB11}, recomputed with a slightly higher numerical
  precision, and the improved 1D LHD structures (dashed blue line) described
  in the text.
\label{fg:f1}}
\end{center}
\end{figure}

The treatment of convection is obviously a genuine difference between the 3D
CO$^{5}$BOLD code and the standard 1D codes. As a consequence, it is difficult
to isolate other possible differences between both codes. This hurdle is
removed by using the 1D LHD code, in which microphysics and radiative transfer
numerical schemes (e.g., the opacity binning) are identical to those included
in the 3D simulations. In Fig.~\ref{fg:f1}, we compare our latest 1D LHD
structures and standard 1D structures, which shows that we were successful in
explaining and eliminating as much as possible the differences between both
codes. Our improvements made to the LHD code are described in this section.

\subsection{Input microphysics}

In Paper I, the CO$^{5}$BOLD and LHD structures were based on different
microphysics \citep{ludwig94} in comparison to the one used in prevalent 1D
white dwarf codes. The former codes rely on pre-computed EOS and opacity
tables as input, and it is fairly straightforward to replace these tables with
ones that are computed with the same microphysics as in standard 1D
models. This is the first step of our improvement of the LHD structures. We
found that the non-ideal EOS typically used in white dwarfs has a negligible
effect on the predicted structures (for $T_{\rm eff} > 6000$~K and $\log
\tau_{\rm R} < 3$) compared to the ideal EOS utilised in Paper I. In the case of the
improved opacities, the effect is mild, and it is shown in Sect.~4.3 that the
corresponding $\log g$ correction is of the order of 0.03 dex.

We now look at the opacity binning approach, which is a known difference
between the LHD and standard white dwarf codes. In Paper I, we used 7
band-averaged opacities to describe the band-integrated radiative transfer,
based on the procedure laid out in \citet{nordlund82,ludwig94} and
\citet{voegler04}. These include two bins dedicated to the Lyman
pseudo-molecular satellites \citep{allard82,allard04}. The 1D models
calculated in \citet{TB11}, on the other hand, employ 1813 carefully chosen
frequencies for the radiative transfer (which is sufficient for the broad
lines in DA stars).

\begin{figure}[!h]
\begin{center}
\includegraphics[bb=18 165 672 652,width=4in]{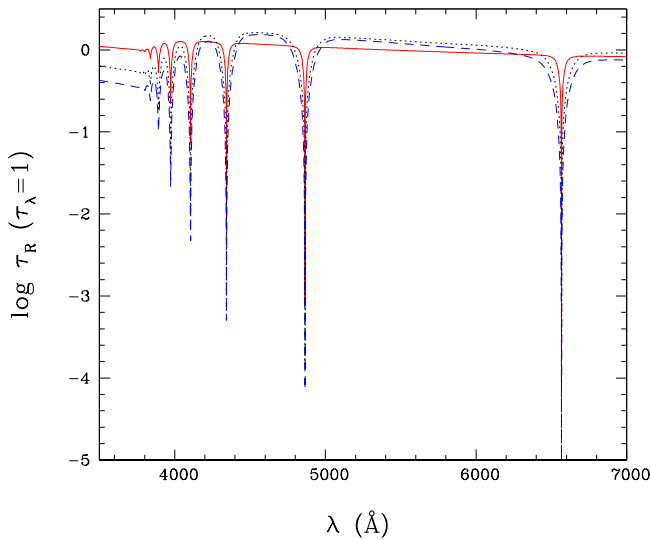}
\caption{Value of the Rosseland optical depth at which the plasma becomes
  optically thin ($\tau_{\lambda} = 1$) at a particular wavelength for 1D model
  atmospheres at 6000 (red, solid), 10,000 (black, dotted) and 12,000~K
  (blue, dashed).
\label{fg:f2}}
\end{center}
\end{figure}

Fig.~\ref{fg:f2} presents the characteristic forming region ($\tau_{\lambda} =
1$) of the emergent flux as a function of the wavelength. We rely on the
Rosseland optical depth ($\tau_{\rm R}$) as the reference scale.
Unsurprisingly, the lines are formed higher in the atmosphere than the
continuum, since the opacity is larger at these wavelengths. We see that the
atmospheric region within $-5 < \log \tau_{\rm R} < 0$ is relevant for the
line formation.

We sort the wavelength-dependent opacities based on the Rosseland optical
depth at which $\tau_\lambda$ = 1. In other words, the boundaries of the
opacity bins could be represented by horizontal lines in Fig.~\ref{fg:f2}. We
use a total of 8 opacity bins in our calculations, by applying thresholds in
$\log \tau_{\rm R}$ given by [$\infty,0.0,-0.5,-1.0,-2.0,-3.0,-4.0,-\infty$]
and adding one bin for the Lyman satellites. We find that it is necessary to
use at least one bin for each dex in $\tau_{\rm R}$ to reproduce the standard
1D calculations in that $\tau_{\rm R}$ region. We also split in two bins the
critical $-1 < \log \tau_{\rm R} < 0$ region to be on the safe side. In all
but two bins a switching between Rosseland and Planck averages is performed at
a band-averaged Rosseland optical depth of 0.35. In the two bins gathering the
largest line opacities, the Rosseland mean opacity is used throughout. We
treat scattering as true absorption. We have verified with the \citet{TB11}
atmosphere code that this approximation has a negligible effect on the
structure of convective DA white dwarfs. Fig.~\ref{fg:f1} shows that we can
reproduce very well the established 1D structures with our opacity
tables. This is an important result, since it was not obvious that we could
reproduce with accuracy the structures of standard 1D models with only a few
opacity bins. We note, in comparison, that heating rates in the Sun are well
reproduced with a 9-bin/12-group scheme \citep{freytag12}.

\subsection{Treatment of convection}

One aspect that could explain differences between the LHD and standard
structures is a difference in the physical parameters that are allowed to
vary. Unfortunately, we discovered such incompatibility which is related to
the MLT convection input parameters used to compute LHD structures. The LHD
code setup in this study relies on the ML2/$\alpha$ = 0.8 parameterisation of
the MLT that is used in the most recent white dwarf models \citep{TB10}, where
the calibration of the mixing-length parameter $\alpha$ results from a
comparison of near-UV and optically determined $T_{\rm eff}$. For
completeness, we summarise in Table~1 the $a, b$ and $c$ parameters that are
used to define the ML2 treatment of the MLT equations \citep[see,
  e.g.,][]{mihalas78,tassoul90,bergeron92b,koester94,ludwig99}.  There is a
fourth parameter that was introduced early in the usage of the MLT theory
\citep{vitense53} to allow for optically thin convective cells \citep[see
  eq. 7.72 of][]{mihalas78}. This parameter is defined as $d$ in the white
dwarf field \citep[see eq. 4 of][]{bergeron92b}, and a $f_4$ parameter,
specified in a different way but recovering the \citet{mihalas78} relation if
$f_4 = 2$, is defined in Table~A1 of \citet{ludwig99}. By changing the value
of $f_4$, the convective efficiency is modified in the transition between the
optically thin and thick regions, which is the region of line
formation. Therefore, this {\it extra} parameter can have a major effect on
the predicted structures and spectra.

 \begin{table}[!h]
 \caption{Adopted parameterisation of MLT}
 \label{tab1}
 \begin{center}
 \begin{tabular}{lll}
\hline
\hline
Param. & Alt. name & Value \\
\hline
$l/H_{\rm p}$ & $\alpha$ & 0.8 \\
$a$ & $-$ & 1 \\
$b$ & $-$ & 2 \\
$c$ & $-$ & 16 \\
$d$ & $\Gamma (1+f_4/\tau_e^2)$ & $\Gamma (1+2/\tau_e^2)$ \\
\hline

\end{tabular} 
\end{center} 
\tablefoot{$l$ is the mixing-length, $H_{\rm p}$ the pressure scale height, $\Gamma$
  the convective efficiency and $\tau_e$ the optical thickness of a convective
  eddy.}
\end{table}

In Paper I, the LHD models were computed with optically thick convective cells
($f_4 = 0$) while in the present work, LHD models are allowed to have
optically thin cells ($f_4 = 2$). As a result the LHD code is now based on an
implementation of the MLT theory that is exactly the same as in commonly used
white dwarf models. The effect on the 3D $\log g$ corrections is discussed in
Sect.~4. We point out that 3D simulations are completely unaffected by the MLT
parameterisation.

\subsection{Numerical precision}

Finally, we had to increase the numerical precision in order to provide the
best match between the LHD and standard structures. We found that at least 200
depth points must be used with the LHD code in order to recover the
\citet{TB11} structures. Furthermore, we find with both codes that the
equation of radiative transfer must be solved for at least five angles. We
stress that our results in terms of the numerical precision of the LHD code
can not easily be transferred to the CO$^{5}$BOLD code, because of the
completely different treatment of convection. Instead, we show in Sect.~4.3
that the sensitivities of the 3D structures to different numerical parameters
is very small and we conclude that our CO$^{5}$BOLD code setup provides the
required high precision.

\section{3D model atmospheres}

We have computed a set of twelve 3D DA model atmospheres with the CO$^{5}$BOLD
code\footnote{We rely on the January 2012 version of the code which is similar
  to the version described in \citet{freytag12}. We utilise some new features
  that we describe in this section.} covering the range $13,000 >$ $T_{\rm
  eff}$ (K) $> 6000$ at $\log g = 8.0$. Table~2 gives a summary of the
properties of these simulations. We have used the new EOS and opacity tables
that were discussed in Sect.~2. We relied upon four different opacity tables,
with the opacity bins sorted based on a 6000, 8000, 10,000 and 12,000~K 1D
reference model atmosphere, respectively. The implementation of the boundary
conditions is described in detail in \citet[][see Sect.~3.2]{freytag12} and
CO$^5$BOLD solar models rely on the same conditions. In brief, the lateral
boundaries are periodic, and the top boundary is open to material flows and
radiation. The bottom layer is open to convective flows (and radiation) in all
but the three hottest simulations, and a zero total mass flux is enforced. We
specify the entropy of the ascending material to obtain approximately the
desired $T_{\rm eff}$ value (derived from the emergent stellar flux). The
entropy of the deep layers is increasing monotonically with $T_{\rm eff}$,
hence there is a unique relation between the entropy at the bottom boundary
and the effective temperature of a simulation. This control of the $T_{\rm
  eff}$ of a model (i.e. the temporal and spatial average of the emergent
radiative flux) is indirect and $T_{\rm eff}$ is not an input
parameter\footnote{For clarity, we use round numbers in the discussion of the
  models in Table~2, but all calculations are done with the exact $T_{\rm
    eff}$ values.}. In contrast, the bottom layer is closed with imposed zero
vertical velocities (but open to radiation) for the three hottest simulations,
in which case we impose the radiative flux according to the diffusion
approximation.

We typically started our simulations by scaling approximately another model
from the grid. As long as the program runs smoothly, the initial structure
does not matter much \citep{freytag12} for space- and time-averaged
quantities. We first computed grey models for 10 seconds, and then switched on
the non-grey radiative transfer for 10 seconds or more. More details about the
time evolution of our simulations is given is Sect.~3.3.

We adopt a grid of $150\times150\times150$ points in the $x,y$ and $z$
directions, where $z$ is used for the vertical direction and points towards
the exterior of the star. Compared to the simulations computed in Paper I, the
vertical number of points was increased from 100 to 150. The geometrical
dimensions are given in Table~2. We fixed the bottom layer at $\log \tau_{\rm
  R} = 3$ for all simulations, well below the photosphere\footnote{We define
  the photosphere as the line-forming regions as opposed to the atmosphere
  which is the full simulation.}. One exception is the third hottest
  model ($T_{\rm eff} \sim 12,000$ K) for which we recomputed a more extended
  simulation with a bottom boundary at $\log \tau_{\rm R} = 3.1$. The bottom
  of the convective zone was too close to the simulation boundary at $\log
  \tau_{\rm R} = 3.0$, which was artificially damping the convective
  velocities by a small amount. Nevertheless, we find that our two
  simulations with an unequal vertical extent show very little differences
  except for the sub-photospheric convective velocities. In all models, the
top boundary reaches a space- and time-averaged value of no more than $\log
\tau_{\rm R} \sim -5$. In Table~2 we also show the number of pressure
  scale heights between the photosphere and the bottom boundary that are
  covered by the simulations. It demonstrates that convective eddies reaching
  the photosphere are unlikely to be impacted by boundary conditions. The
grid spacing in the $z$ direction is non-equidistant. The horizontal
geometrical dimensions are further in discussed in Sect.~3.4.

 \begin{table*}[!]
 \caption{3D model atmospheres}
 \label{tab2}
 \begin{center}
 \begin{tabular}{lccccc}
\hline
\hline
$T_{\rm eff}$ & $\log g$ & $ x\times y\times z$ & $\ln(P_{\rm bot}/P_{\rm phot})$\tablefootmark{a} & Time\tablefootmark{b} &  3D
$\log g$ corr. \\
(K) & & (km)$\times$(km)$\times$(km) &  & (s) & \\
\hline
5997 & 8.0 & 1.19$\times$1.19$\times$0.34 & 5.0 & 100 &  0.00 \\
7012 & 8.0 &  1.19$\times$1.19$\times$0.38 & 4.2 & 60 & -0.07 \\
8032 & 8.0 &  1.40$\times$1.40$\times$0.43 & 3.7 & 60 & -0.16 \\
9035 & 8.0 &  1.58$\times$1.58$\times$0.51 & 3.3 & 10 & -0.26 \\
9520 & 8.0 &  2.06$\times$2.06$\times$0.62 & 3.8 & 10 & -0.25 \\
10018 & 8.0 & 2.10$\times$2.10$\times$0.76 & 4.3 & 10 & -0.24 \\
10530 & 8.0 & 2.24$\times$2.24$\times$0.88 & 3.6 & 10 & -0.22 \\
11004 & 8.0 & 2.68$\times$2.68$\times$1.19 & 4.0 & 10 & -0.15 \\
11531 & 8.0 & 3.65$\times$3.65$\times$1.69 & 3.6 & 10 & -0.11 \\
12022 & 8.0 & 7.45$\times$7.45$\times$4.06 & 2.0 & 10 & -0.07 \\
12505 & 8.0 & 7.45$\times$7.45$\times$4.08 & 1.9 & 10 & -0.09 \\
12999 & 8.0 & 7.45$\times$7.45$\times$4.41 & 1.8 & 10 & -0.09 \\
\hline

\end{tabular} 
\end{center} 
\tablefoottext{a}{$P_{\rm bot}$ is the pressure at the bottom layer and
  $P_{\rm phot}$ the pressure at $\tau_{\rm R} = 2/3$.}
\tablefoottext{b}{Total stellar time. For the three cooler models, the total
  time includes the initial 2D run as described in Sect.~3.3.}
\end{table*}

\begin{figure*}[]
\captionsetup[subfigure]{labelformat=empty}
\begin{center}
\subfloat[]{
\includegraphics[bb=100 400 572 575,width=4.5in]{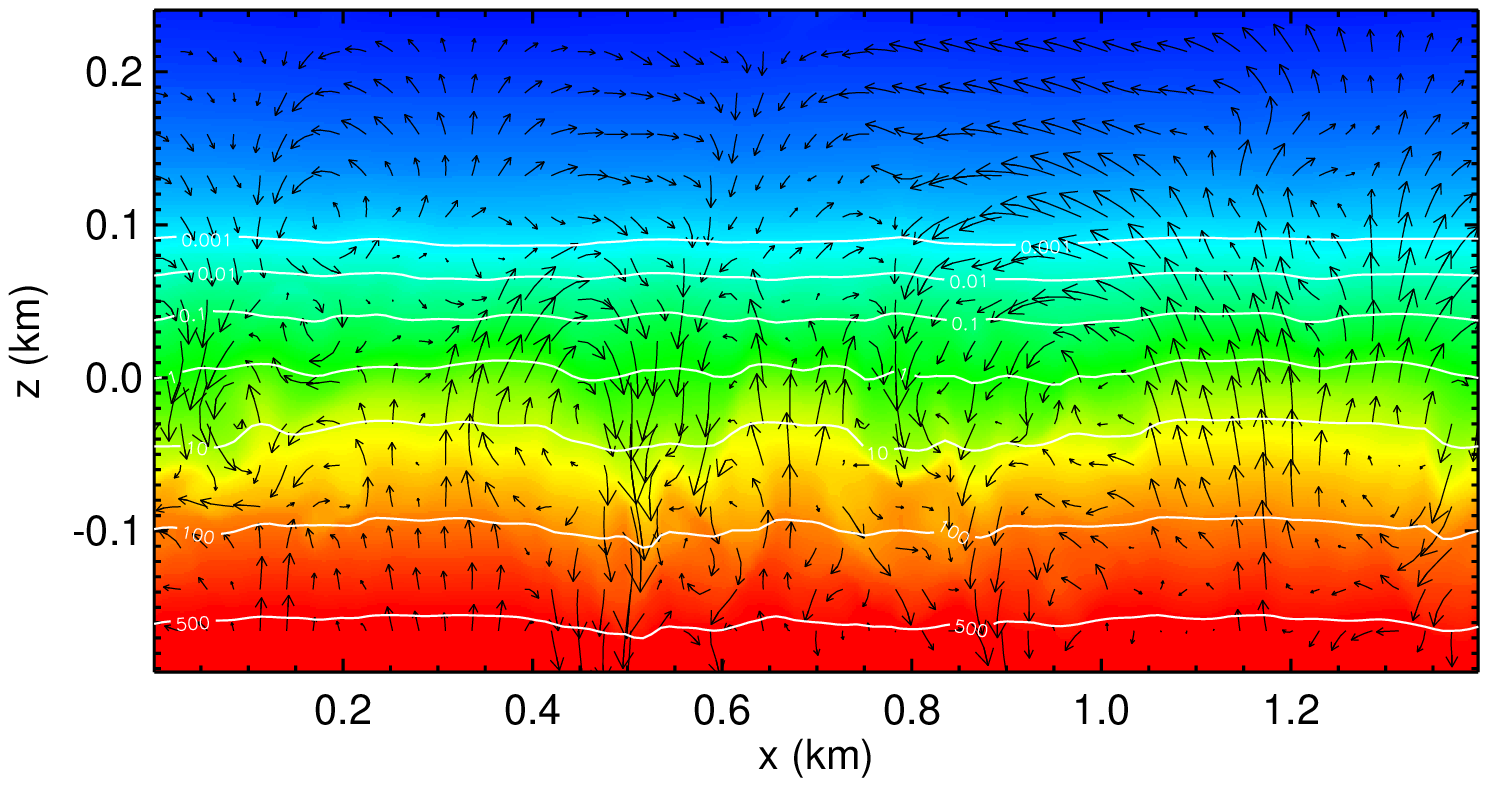}}
\subfloat[]{
\includegraphics[bb=130 400 572 675,width=3.75in]{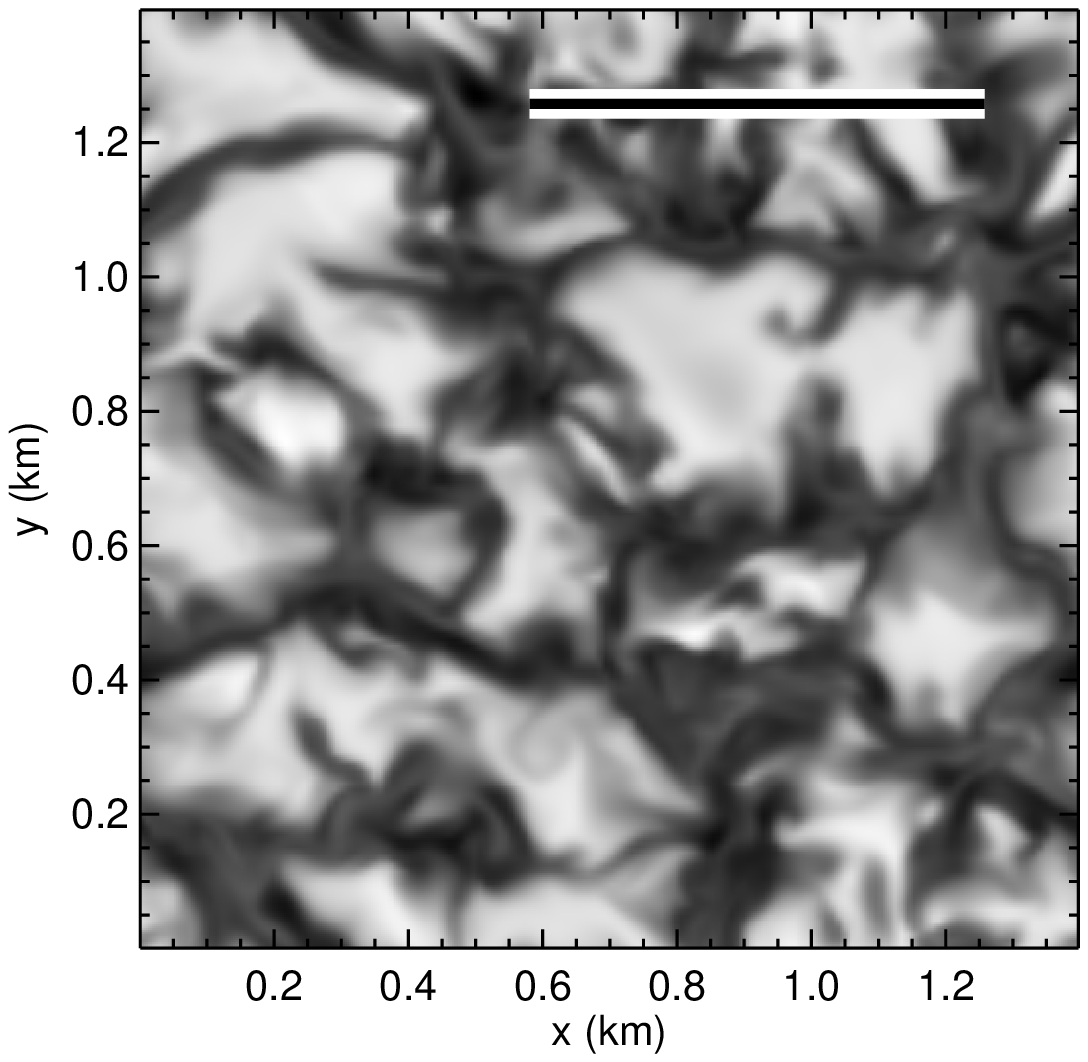}}
\caption{Snapshot of the 3D white dwarf simulation at $T_{\rm eff} \sim 8000$~K and
  $\log g = 8$. {\it Left:} Temperature structure for a slice in the
  horizontal-vertical $xz$ plane through a box with coordinates $x,y,z$ (in
  km). The temperature is colour coded from 14,000 (red) to 3000~K (blue). The
  arrows represent relative convective velocities, while thick lines
  correspond to contours of constant Rosseland optical depth, with values
  given in the figure. {\it Right:} Emergent bolometric intensity at the top
  of the horizontal $xy$ plane. The RMS intensity contrast with respect to the
  mean intensity is 7.6\%. The length of the bar in the top right is 10
  times the pressure scale height at $\tau_{\rm R} = 2/3$.
\label{fg:f3}}
\end{center}
\end{figure*}

\begin{figure*}[]
\captionsetup[subfigure]{labelformat=empty}
\begin{center}
\subfloat[]{
\includegraphics[bb=100 400 572 575,width=4.5in]{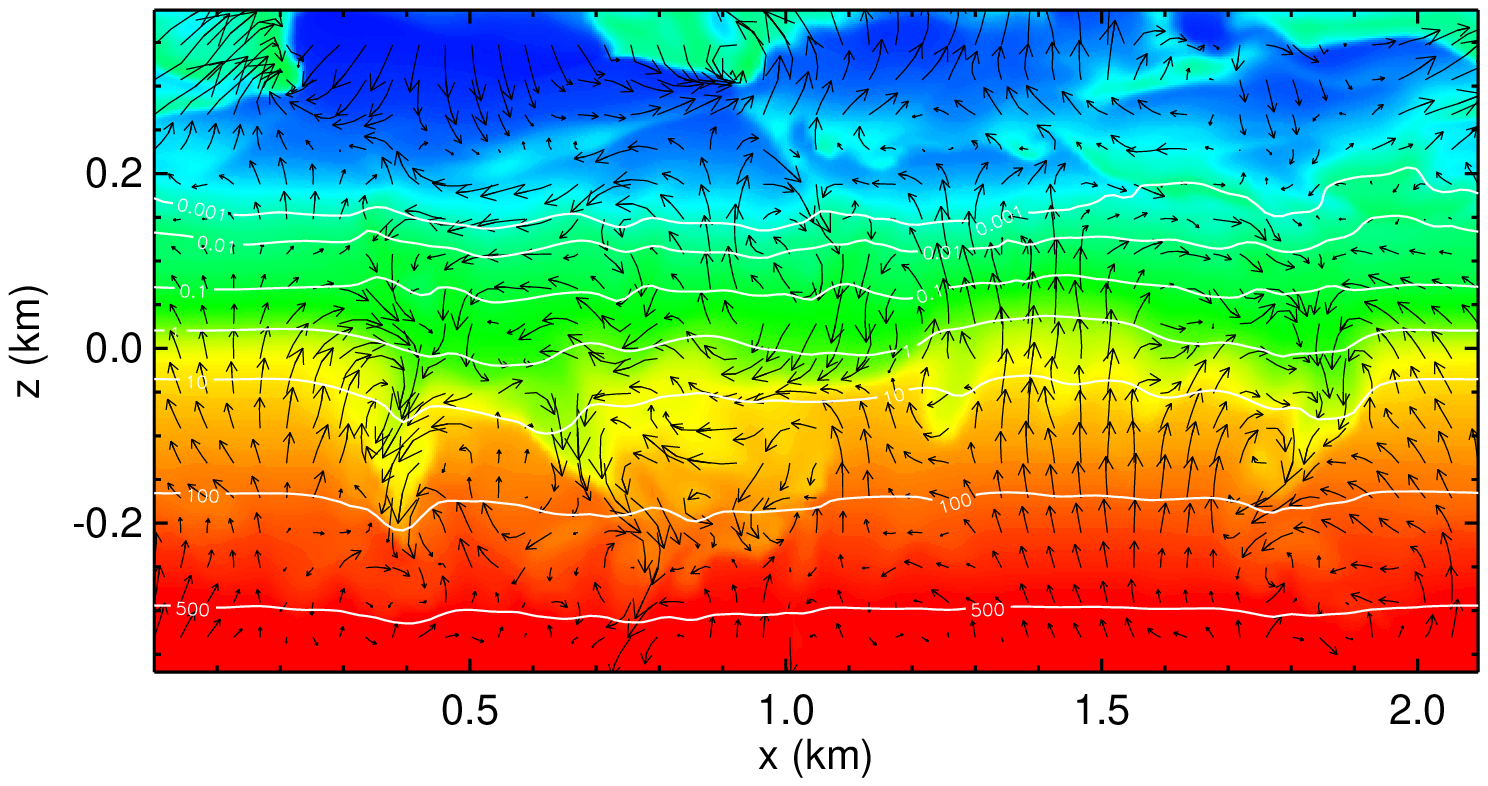}}
\subfloat[]{
\includegraphics[bb=130 400 572 675,width=3.75in]{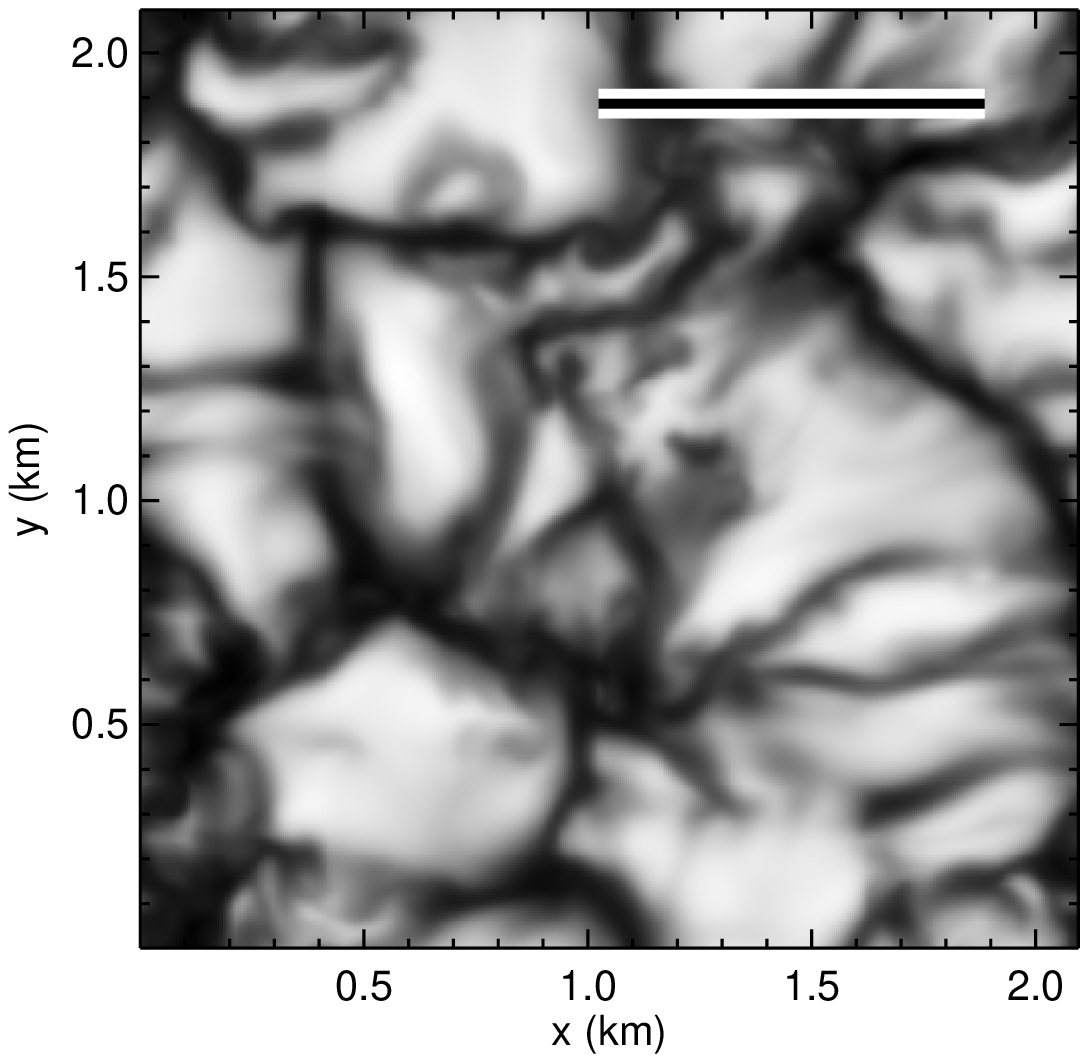}}
\caption{Similar to Fig.~\ref{fg:f3} but for the simulation at $T_{\rm eff} \sim
  10,000$~K. {\it Left:} The temperature is colour coded from 17,000 (red) to
  5000~K (blue). {\it Right:} The RMS intensity contrast is 14.4\%.
\label{fg:f4}}
\end{center}
\end{figure*}

\begin{figure*}[]
\captionsetup[subfigure]{labelformat=empty}
\begin{center}
\subfloat[]{
\includegraphics[bb=100 400 572 575,width=4.5in]{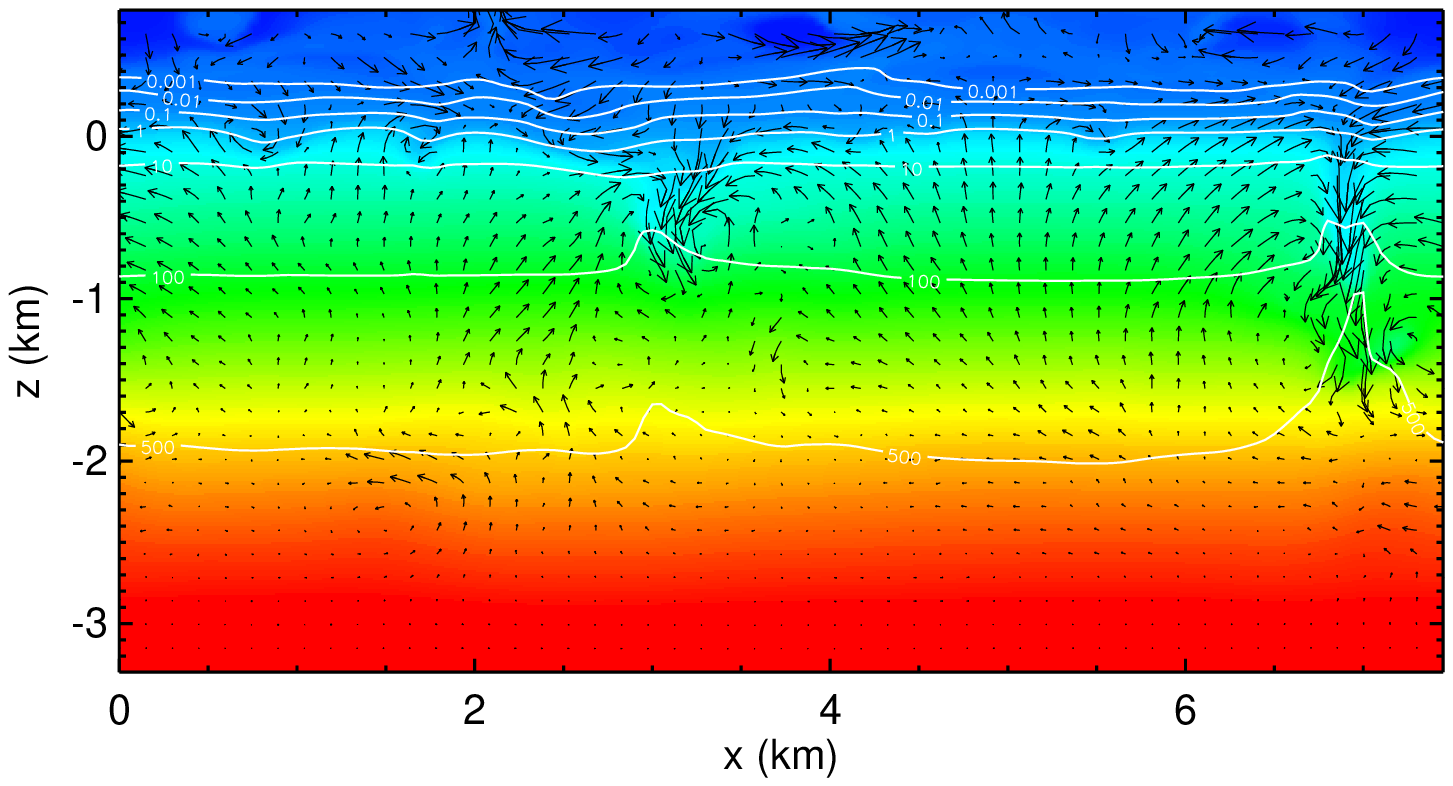}}
\subfloat[]{
\includegraphics[bb=130 400 572 675,width=3.75in]{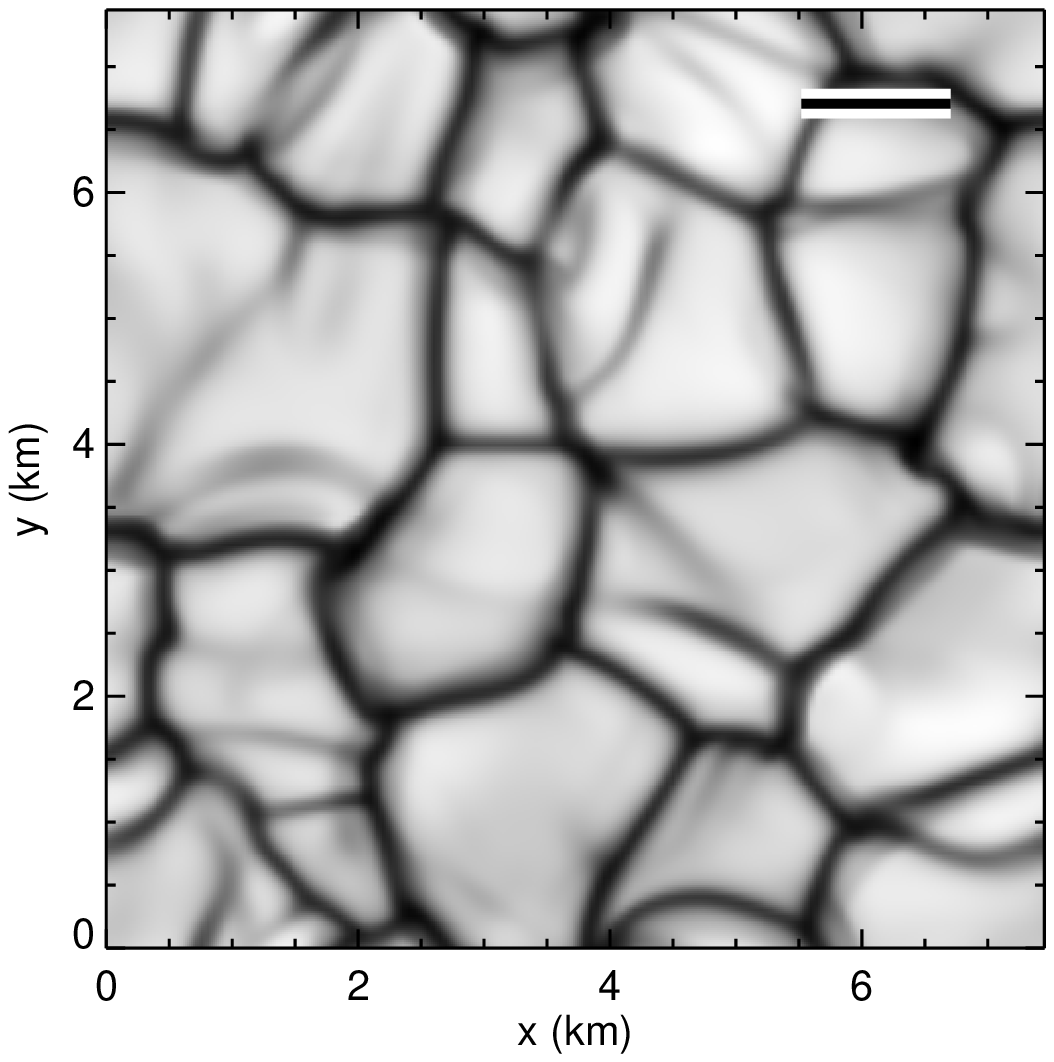}}
\caption{Similar to Fig.~\ref{fg:f3} but for the simulation at $T_{\rm eff} \sim
  12,000$~K. {\it Left:} The temperature is colour coded from 60,000 (red) to
  7000~K (blue). {\it Right:} The RMS intensity contrast is 18.8\%.
\label{fg:f5}}
\end{center}
\end{figure*}

In terms of solving the hydrodynamical equations, the van Leer slope limiter
was used in Paper I, but we have now switched to a less dissipative
2nd-order reconstruction method. Furthermore, regarding the time integration
scheme, we now adopt the corner-transport upwind (CTU) method \citep{colella90}.

It is well known that the flows in stellar atmospheres are characterized by
very large Reynolds numbers (Re $> 10^{10}$), and white dwarfs are not an
exception. This implies that the flows are highly turbulent, and that the
turbulent kinetic energy is then dissipated into heat at the Komolgorov
microscale ($d \sim H_{\rm p} {\rm Re}^{-3/4}$), i.e. on scales much smaller
than the resolution of our grid. As a consequence, only the largest flow
structures are resolved, and the small-scale kinetic energy is dissipated at
the grid scale, either by the Roe solver itself or by an additional artificial
tensor viscosity. The granules are the large flow structures that contribute
to the global convective energy exchanges in the atmospheres, and as long as
they are well resolved, the numerical treatment of the energy dissipation at
smaller scales has very little effect on the predicted structures. Hence,
viscosity is not a free parameter like, for instance, the mixing-length
parameters. In Paper I, artificial viscosity was added, but with our new setup
of the numerical schemes, we could compute all simulations without this explicit
viscosity. This allows resolving a bit further the flows for which the spatial
scale is close to the resolution of the box. Finally, we note that the
 radiation pressure is negligible for these models.

Figs. \ref{fg:f3} to \ref{fg:f5} present the final snapshots for three
  of our simulations at $T_{\rm eff}$ $\sim$ 8000, 10,000 and 12,000~K. The
  first series of plots (left panels) present the atmospheric structures in
  terms of temperature contours and velocity fields as a function of the
  geometrical depth ($z$) and one of the horizontal direction ($x$). We also
  show surfaces of constant Rosseland optical depth over which the mean
  structures are computed throughout this study.

In all snapshots, significant velocities are observed in the upper layers
($\tau_{\rm R} \lesssim 10^{-3}$) even though these regions are stable to
convection. The presence of oscillating waves is the main explanation for this
behaviour, and it is difficult to differentiate these waves from the
exponentially decaying convective flows (overshoot layer), which transports
energy outside of the convective zone (see Sect.~3.2). Another obvious
observation is that the velocity field is significantly different for the
hotter 12,000 K simulation. It is seen that the downdrafts are concentrated in
narrow lanes in comparison to the cooler models. Furthermore,
roughly the lower third of the 12,000~K simulation, in terms of geometrical
depth, is stable to convection. However, large downdrafts, such as the one on
the far right of Fig. \ref{fg:f5}, reach high $\tau_{\rm R}$ values. The
root-mean-square (RMS) vertical velocities actually remain well above zero up
to $\tau_{\rm R} \sim 10^3$ because of these downdrafts.
  
The second series of plots (right panels) present the emergent frequency
integrated (bolometric) intensity. It is shown that the granulation size is
changing significantly over the range of $T_{\rm eff}$. Once again, the
12,000~K simulation stands out with narrower cool downdrafts and smoother hot
cells. The size, shape and contrast of the granulation patterns will be
discussed in more details in Sect.~3.4.

\subsection{Mean structures}

We have shown in Sect.~2 that by using the proper EOS and opacity
tables, it is possible to reproduce standard white dwarf model atmospheres
with the LHD code. Given that CO$^{5}$BOLD employs the same microphysics and
radiative transfer schemes as LHD, we can conclude that CO$^{5}$BOLD reaches a
similar {\it precision}, and possibly a better {\it accuracy} due the improved
convection treatment, in comparison to the current 1D models. We now compare
directly the results from the 3D simulations with the 1D models of \citet{TB11}.

It was demonstrated in Paper I (Fig.~4) that in the case of DA white dwarfs,
3D spectral synthesis produces the same result as computing one $\langle {\rm
  3D}\rangle$ spectrum from a single $\langle {\rm 3D}\rangle$ structure,
which is the temporal and spatial average of a 3D simulation over surfaces of
constant Rosseland optical depth. The mean temperature and pressure
structures, and the atmospheric parameters ($T_{\rm eff}$ and $\log g$), are
the quantities necessary to produce model spectra from the 3D simulations,
which is the objective of Sect.~4. We derived mean $T$ and $P$ values from the
average of $T^4$ and $P$ over surfaces of constant $\tau_{\rm R}$. We selected 12
random snapshots in the last 2.5 seconds of the simulations to make the
temporal average. We determined the $T_{\rm eff}$ values identified in Table~2
from the mean emergent flux of the 12 snapshots selected for the $\tau_{\rm R}$
average. The RMS variation of $T_{\rm eff}$ with time is always much less than
1\%. In contrast, $\log g$ is a fixed quantity in the hydrodynamical
equations.

\begin{figure}[!h]
\begin{center}
\includegraphics[bb=18 165 672 652,width=4in]{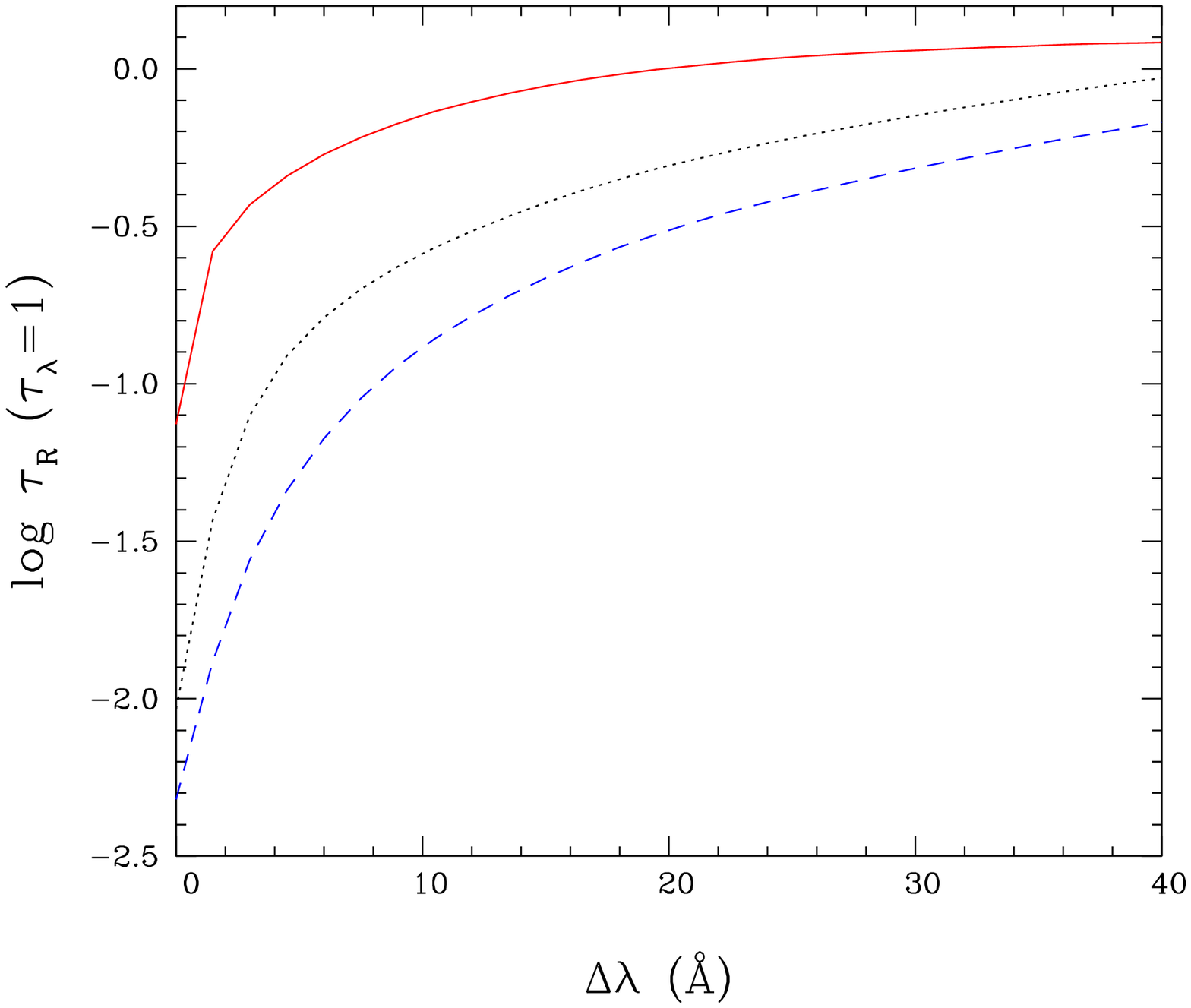}
\caption{Similar to Fig.~\ref{fg:f2} but in the wavelength range, with respect
  to the line center, of the red wing of H$\delta$.
\label{fg:f6}}
\end{center}
\end{figure}

To understand the differences between $\langle {\rm 3D}\rangle$ and 1D models,
it is useful to look at other mean quantities. Therefore, we computed the mean
entropy and RMS velocities at constant optical depth, and different energy
fluxes described below at constant geometrical depth. Fig.~\ref{fg:f6}
illustrates that the characteristic depth of the formation of H$\delta$, taken
as a typical line, is within the range $-2 < \log \tau_{\rm R} < 0$. Hence this
region of the photosphere is the most critical for the predicted spectra.

\begin{figure}[!h]
\begin{center}
\includegraphics[bb=18 155 662 712,width=4in]{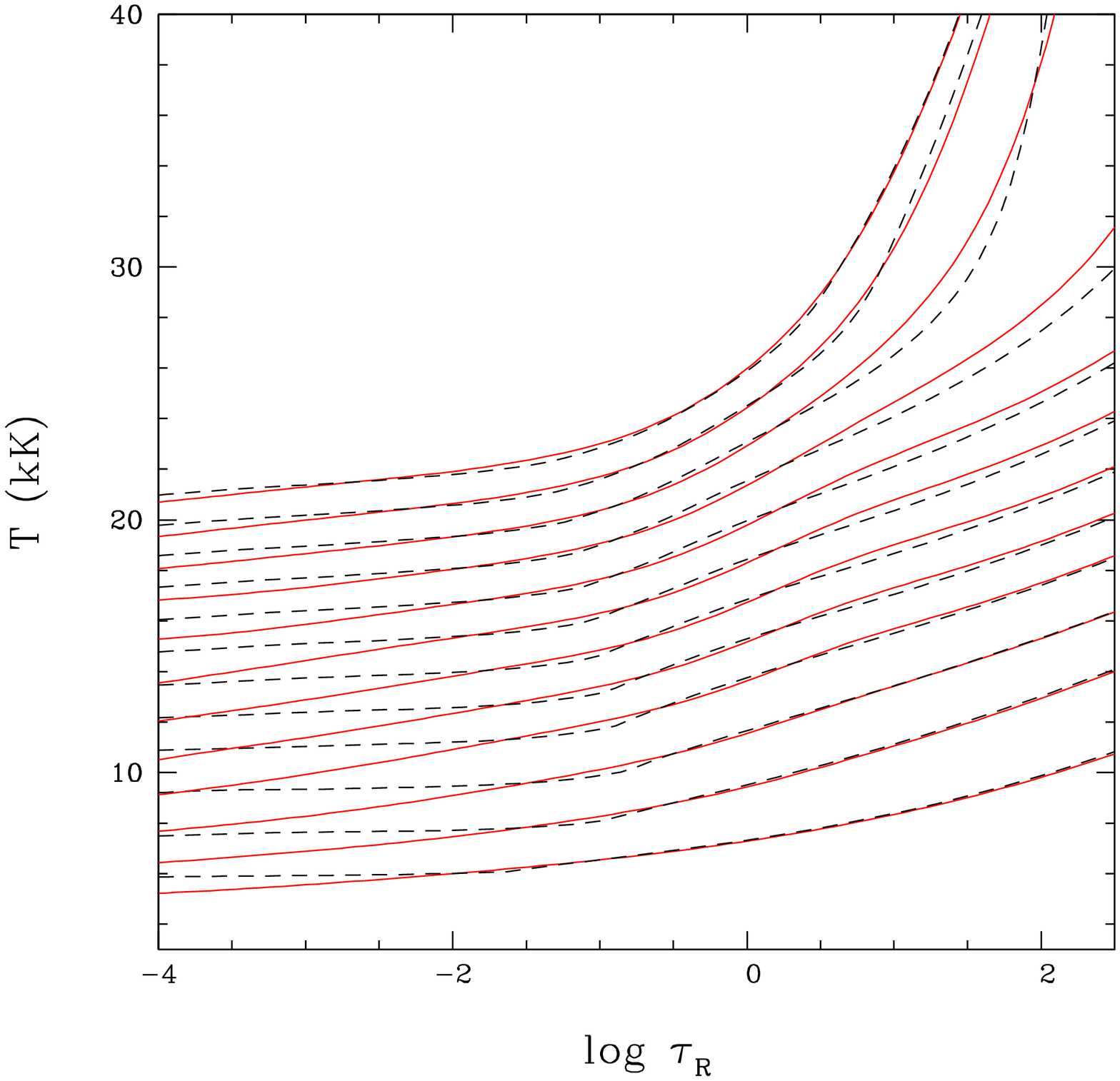}
\caption{Temperature structures versus log $\tau_{\rm R}$ for $\langle {\rm
    3D}\rangle$ (red, solid) and 1D (black, dashed) model atmospheres. The
  temperature scale is correct for the 6000~K model (bottom curve), but other
  structures are shifted by 1 kK relative to each others for clarity.
\label{fg:f7}}
\end{center}
\end{figure}

\begin{figure}[!h]
\begin{center}
\includegraphics[bb=18 140 662 712,width=4in]{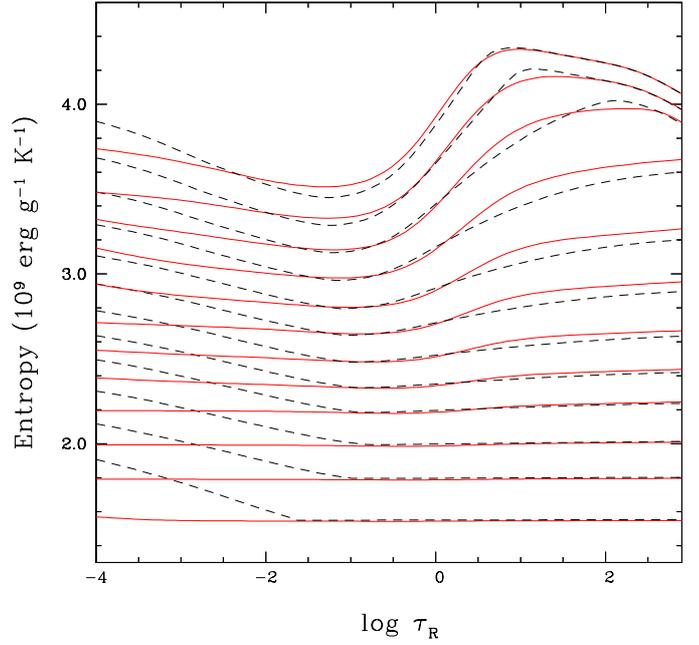}
\caption{Entropy as a function of $\log \tau_{\rm R}$ for our sequence of $\langle
  {\rm 3D}\rangle$ (red, solid) and 1D models (black, dashed), with the
  coolest model at the bottom, and the hottest model at the top. The entropy
  scale is correct for the 6000~K model, but other structures were shifted by
  0.1 unit for clarity.
\label{fg:f8}}
\end{center}
\end{figure}

In Figs. \ref{fg:f7} and \ref{fg:f8}, we compare the $\langle {\rm
  3D}\rangle$ and 1D structures in terms of the temperature and entropy,
respectively. The 1D models are drawn from the grid of \citet{TB11} with the
ML2/$\alpha$ = 0.8 parameterisation of the MLT. Compared to the temperature,
the entropy values are more sensitive to differences between $\langle {\rm
  3D}\rangle$ and 1D structures. Furthermore, the presence of a negative or
positive entropy gradient (as a function of $\tau_{\rm R}$) shows whether a
layer is stable or unstable to convection, respectively. It is observed that
in stability terms, the top of the convective zone, and even the bottom of the
zone for the three hottest structures, are at similar optical depths in $\langle
{\rm 3D}\rangle$ and 1D models. In the upper part of the atmospheres, the
entropy gradient (in absolute value) and the temperatures are always much
smaller in $\langle {\rm 3D}\rangle$ structures, which is the result of convective
overshoot. This overshoot layer is able to cool the upper layers because of a
weak radiative coupling (see Sect.~3.3). In comparison, the 1D models are
stable to convection and in radiative equilibrium.

\begin{figure}[!h]
\begin{center}
\includegraphics[bb=18 140 662 712,width=4in]{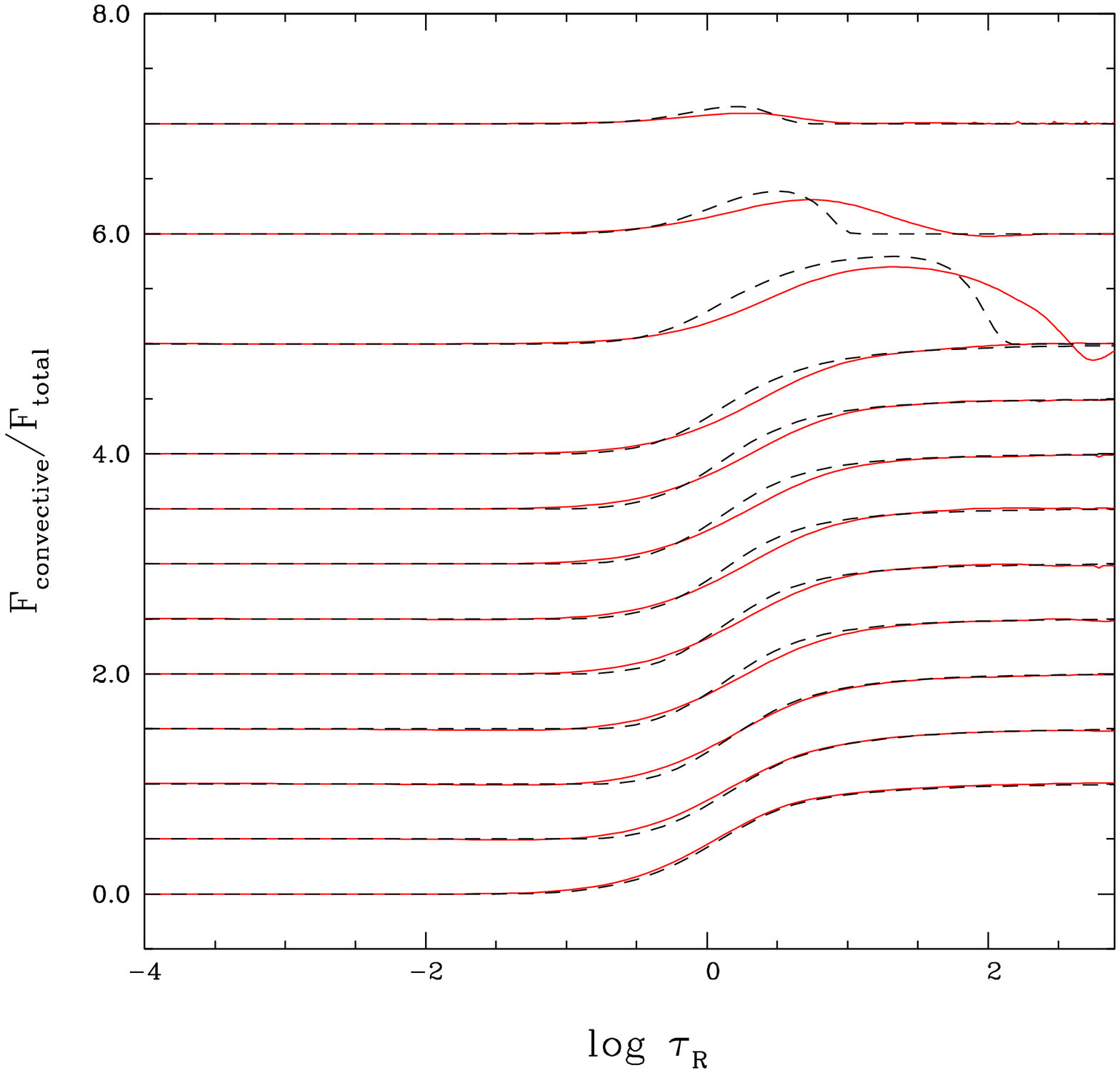}
\caption{Ratio of the convective to the total energy flux as a function of
  $\log \tau_{\rm R}$ for our sequence of $\langle {\rm 3D}\rangle$ (red, solid) and
  1D models (black, dashed), with the coolest model at the bottom, and the
  hottest model at the top. The ratio is exact for the 6000~K model, but other
  structures were shifted by 0.5 up to 11,500~K, and by 1.0 above that
  temperature, for clarity.
\label{fg:f9}}
\end{center}
\end{figure}

In Fig.~\ref{fg:f9}, we present the ratio of the convective flux to the total
energy flux as a function of optical depth. The convective flux is the sum of
the enthalpy, gravitational energy and kinetic energy fluxes. The shape of the
convective zones as described by the convective flux is also similar between
$\langle {\rm 3D}\rangle$ and 1D models. One reason is that the overshoot in
the upper layers acts in optically thin regions, and transports very little
flux. For the three hottest models, however, the convective zone is
systematically deeper for the $\langle {\rm 3D}\rangle$ models in part because
an overshoot layer is present at the bottom of the convection zone where it is
able to transport energy. This is in qualitative agreement with the results
from pulsation studies in which a slightly more efficient MLT parameterisation
(ML2/$\alpha$ = 1.0) is used to describe the bottom of the convective zones
\citep{fontaine08}. We note that larger than average differences in the
$\langle {\rm 3D}\rangle$ and 1D temperature structures at $\log \tau_{\rm R}
>1$ for the hot models are due to the transition from convective to radiative
layers. In general, the transition region from convective to radiative flux
transport is smoother in $\langle {\rm 3D}\rangle$ structures according to
Fig.~\ref{fg:f9}, which explains why $\langle {\rm 3D}\rangle$ structures
have smoother temperature gradients in the photosphere as reported by
Fig.~\ref{fg:f7}.

Interestingly, the $\langle {\rm 3D}\rangle$ and 1D models are the most
similar at the boundaries of our computed sequence and the maximum
differences between 1D and $\langle {\rm 3D}\rangle$ models in the photosphere
are detected in the middle of our sequence at $T_{\rm eff} \sim
9000-10,000$~K. At $T_{\rm eff} \sim$ 6000~K, convection is almost fully
adiabatic everywhere in the atmosphere, and since 1D and 3D models are based
on the same input microphysics, it is expected that the adiabatic gradient,
hence the structures, will be the same. For the hottest model ($T_{\rm eff}
\sim$ 13,000~K), the maximum convective flux becomes small, and the atmosphere
is leading towards the fully radiative regime. Nevertheless, it is
  remarkable that the entropy gradients are still substantially different in
  the upper layers, which suggests that the convective overshoot is still
  significant at this temperature. However, part of the differences between
  the 1D and 3D $T_{\rm eff}$ structures may also be caused by the distinct
  treatment of the radiative transfer.

\subsection{Convective velocities}

\begin{figure}[!h]
\begin{center}
\includegraphics[bb=18 140 662 712,width=4in]{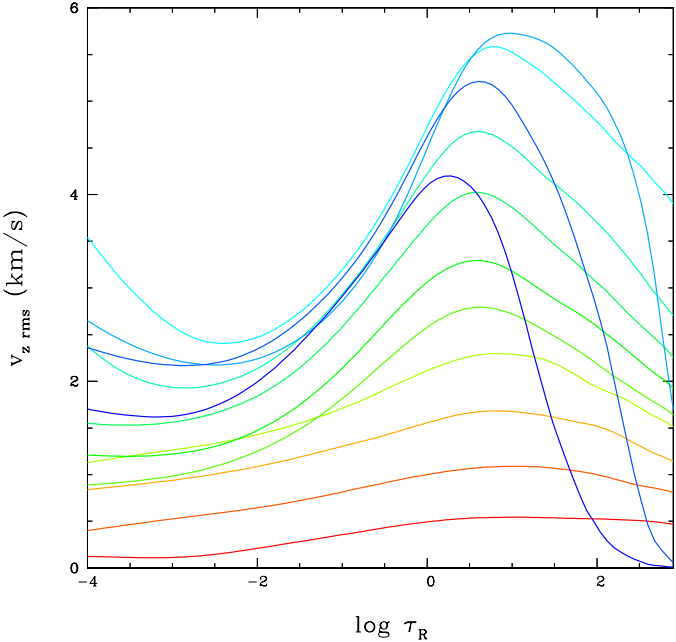}
\caption{Vertical RMS velocity $v_z$ as a function of $\log \tau_{\rm R}$ for our
  sequence of $\langle {\rm 3D}\rangle$ models from
6000 (lowest, red curve) over 10,000 (top, light blue curve) 
to 13,000~K (dark blue curve).
\label{fg:f10}}
\end{center}
\end{figure}

\begin{figure}[!h]
\begin{center}
\includegraphics[bb=18 140 662 712,width=4in]{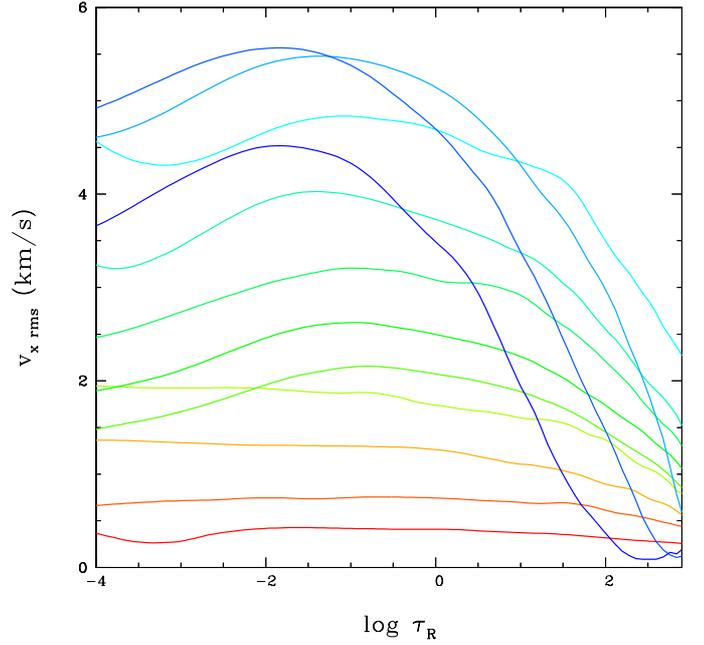}
\caption{Horizontal RMS velocity $v_x$ as a function of $\log \tau_{\rm R}$ for
  our sequence of $\langle {\rm 3D}\rangle$ models from 6000 (lowest, red curve) to 13,000
 ~K (dark blue).
\label{fg:f11}}
\end{center}
\end{figure}

Another relevant aspect of the 3D simulations are the convective
velocities. We present the RMS vertical and horizontal velocities in
Figs. \ref{fg:f10} and \ref{fg:f11}. Due to the symmetry of the simulations,
$x$ and $y$ velocities are nearly identical and we only show the former
component. Our sequence of models exhibits smooth changes of the velocities
between adjacent $T_{\rm eff}$ values, which further confirms that our
structures are well converged. The vertical velocities in Fig.~\ref{fg:f10}
are the most interesting since we can relate them to the convective energy
flux. The maximum vertical velocities are observed in the 11,500 and 12,000~K
simulations and the values decrease in a regular way in cooler and hotter
models. The velocities in the coolest computation at 6000~K are only a small
fraction of the maximum velocities observed in our sequence, which illustrates
the fact that only mild convection is needed to transport the small total
energy flux at 6000~K.

The overshoot layer at the bottom of the convective zones is clearly seen in
terms of the convective velocities. For the three hottest simulations, the
velocities remain clearly above zero even at the bottom of the simulations
($\log \tau_{\rm R} \sim 3$) while Fig.~\ref{fg:f8} shows, from the change of
sign in the entropy gradient, that the layers become stable to convection at
$\log \tau_{\rm R} \sim $ 1.1, 1.5 and 2.2 for the 13,000, 12,500 and 12,000~K
simulations, respectively. Clearly, the overshoot layers extend, in $\log
\tau_{\rm R}$, at least 1 dex and up to 2 dex below the region unstable to
convection. However, Fig.~\ref{fg:f9} illustrates that the overshoot layers
transport very little convective energy flux except in the vicinity of
  the convective zone. Clearly, the relevance of the overshooting depends on
the physical processes that are studied. For instance, we believe that in
diffusion or convective mixing studies, a proper account of the overshoot
layer is essential, as also concluded by \citet{freytag96} from 2D
simulations.

Figs. \ref{fg:f10} and \ref{fg:f11} would also suggest large overshoot layers
above the convective zone, but we must be very cautious about this
interpretation. Indeed, p-mode oscillations, with periods that are related to
the geometry of the computational box (and boundary conditions), occur in all
of our calculations. The effect of these oscillations is largely removed from
all the mean quantities by the temporal average, except in the case of RMS
velocities. The amplitude of the p-mode oscillations is larger in the upper
layers, where the pressure restoring force is lower ($v \propto P^{-1/2}$ for
undamped oscillations). In Fig.~\ref{fg:f10}, the increase of the vertical
velocities at $\log \tau_{\rm R} < -2$ is caused by these
oscillations. Methods have been proposed to filter these oscillations
\citep{ludwig02} in order to look at the exponential overshoot more
clearly. However, for the purpose of this work, the effect of overshooting
into the upper layers can be more easily seen from the temperature and entropy
structures.

The turbulent pressure $\langle \rho v_z^2 \rangle$ is a significant fraction
of the local gas pressure in convective white dwarfs. This fraction reaches
values of $\sim$10$\%$ for the hottest simulations of our grid. The turbulent
pressure derived from the RHD simulations illustrates the deviation from
hydrostatic equilibrium. A proper account of the turbulent pressure in
hydrostatic 1D models might help to improve the agreement between 1D and
$\langle {\rm 3D}\rangle$ structures.

\subsection{Characteristic time scales}

We note that three of our simulations ($12,500 > T_{\rm eff}$ (K) $> 11,500$)
are within the ZZ Ceti instability strip \citep{gianninas11}. It would be
adequate to rely on our 3D structures as input for asteroseismic studies
although the coolest ZZ Ceti star in our grid (11,500 K) is still convective
at the bottom of the simulation. One option would be to compute RHD
simulations that reach the bottom of the convective zone for all pulsating
white dwarfs and then combine the RHD results with 1D interior
structures. Such an application has been performed once by \citet{gautschy96}
and it would be appropriate to update this analysis with our improved and less
dissipative 3D calculations. Alternatively, \citet{ludwig99} have shown that
it is generally possible to match a 3D structure including a convective bottom
layer with an unique 1D structure model by relying on the asymptotic value of
the spatially resolved entropy.

The study of time-resolved white dwarf model atmospheres with hydrodynamical
simulations is unparalleled by 1D models. We note that this time resolution is
of interest for asteroseismic studies. It has been understood long ago that
convective processes were fairly rapid in ZZ Ceti white dwarfs compared to the
typically observed pulsation periods in the range of 100-1000 seconds
\citep{fontaine08}. As a consequence, non-adiabatic pulsation codes have
adopted the concept of instantaneous reaction of the convection to the
pulsations \citep{brassard97}. The impact of time-resolved convection on
asteroseismology of ZZ Ceti stars has recently been studied by
\citet{grootel12} using perturbations in the mixing-length equations. However,
they show that the differences are small compared to the instantaneous
reaction approximation.

\begin{figure}[!h]
\begin{center}
\includegraphics[bb=18 135 672 722,width=4.3in]{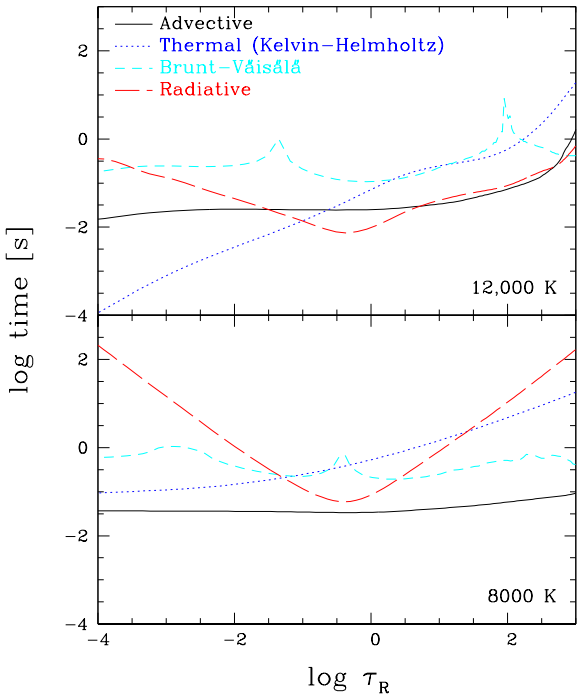}
\caption{Timescales for the $\langle {\rm 3D}\rangle$ structure of a 12,000
  (top panel) and 8000~K (bottom panel) DA white dwarf as a function of log
  $\tau_{\rm R}$. The different timescales are identified in the legend, and
  correspond to the advective timescale (black, solid), the Kevin-Helmholtz
  timescale (blue, dotted), the Brunt-V\"ais\"al\"a timescale (cyan, short
  dashed) and the radiative timescale (red, long dashed).
\label{fg:f12}}
\end{center}
\end{figure}

In Fig.~\ref{fg:f12}, we present four characteristic timescales as a function
of the optical depth in the atmosphere of a 12,000 and 8000~K white
dwarf. The timescales have been evaluated under simplifying assumptions, and
should therefore be taken as order-of-magnitude estimates only. The
timescales are defined as follow \citep[see also][]{ludwig02,freytag12}:

\begin{equation}
t_{\rm BV} = \frac{2 \pi}{\sqrt{\delta_T \vert \nabla_{\rm ad} - \nabla \vert g/H_{\rm p}}}
\end{equation}

\begin{equation}
t_{\rm adv} = \frac{H_{\rm p}}{v_c}
\end{equation}

\begin{equation}
t_{\rm KH} = \frac{P c_{p} T}{g \sigma T_{\rm eff}^4}
\end{equation}

\begin{equation}
t_{\rm rad} = \frac{\rho c_p H_{\rm p}}{16 \sigma \overline{\tau} T^3} \hspace{3mm} {\rm
  where} \hspace{1mm}
\overline{\tau} = \sum_{i}^{\rm opac.~bins} w_i \frac{\kappa_i \rho H_{\rm p}}{2+(\kappa_i \rho H_{\rm p})^2}
\end{equation}

\noindent where $\delta_T \equiv - \left( \frac{\partial {\rm ln}
  \rho}{\partial {\rm ln} T} \right)_P$ denotes the thermal expansion
coefficient at constant pressure, $\nabla_{\rm ad}$ the adiabatic gradient,
$\nabla$ the actual temperature gradient, $v_c$ the RMS convective velocity,
$c_{p}$ the specific heat per gram, $\sigma$ the Stefan-Boltzmann
constant. $\overline \tau$ is the characteristic optical depth of a
disturbance of a size $H_{\rm p}$, with $\kappa_i$ the mean bin opacity per gram,
and $w_i$ the weight of the opacity bin \citep[see eq. A.4 of][]{ludwig12}.

First of all, the Brunt-V\"ais\"al\"a timescale ($t_{\rm BV}$) is related to
the period of the oscillations due to the buoyancy force in regions stable to
convection and the convective growth time for convectively unstable
regions. The {\it spikes} observed for this quantity are related to the
boundaries of the convective layers. The advective timescale ($t_{\rm adv}$),
or the turnover timescale inside the convection zone, is related to the
characteristic time for a convective mass element to move a distance of one
pressure scale height. This timescale varies with depth from 10$^{-2}$ to 1
second and it does not change significantly between the 8000 and 12,000~K
simulations.

The Kelvin-Helmholtz timescale ($t_{\rm KH}$) is the thermal energy content
(per unit area) divided by the total energy flux and describes the thermal
relaxation (trough radiative diffusion). This timescale reaches values of the
order of 15 seconds for the deeper layers of both the 8000 and 12,000~K
models. However, in simulations below 12,000~K, the bottom layer is convective, and
close to $100\%$ of the flux is transported by convection. When the convective
flux is much larger than the radiative flux at the bottom layer, the structure
of the atmosphere is expected to adjust on the advective timescale. For
the three hottest simulations with a radiative bottom layer, we have verified that
in practice, the total flux is stable after a few seconds of computation in all
layers, perhaps due to the presence of convective overshoot and a correct
initial guess of the structure of the deepest layers.

The radiative timescale ($t_{\rm rad}$) refers to the characteristic time for
the decay of local temperature perturbations through radiative transfer. The
radiative timescale is of the order of the advective timescale at 12,000~K,
which is also the case for the 8000~K model, but only in the photosphere. The
P\'eclet number in the photosphere, which is the ratio of the radiative
timescale to the advective timescale, is changing from a value above to below
unity between the cooler and hotter simulations, respectively. This implies that the
evolution of convective cells in the photosphere will be largely influenced by
radiation in the hot models in our grid, whereas for the cool models,
convective cells are more directly governed by quasi-adiabatic expansion and
compression. In the convective zone of the 12,000~K simulation, characteristic of a
ZZ Ceti star, all timescales are lower than one second, which confirms that
the reaction of the convective zone to changing conditions in the radiative
layer just below will be rather instantaneous on the timescales of the
pulsations.

In the 8000~K model, the radiative timescale is rather large at the top and
bottom layers. At the bottom layer, energy is transported by convection and
the large radiative timescale simply implies that the energy in the convective
cells will hardly be lost by radiation. The upper layers above $\log \tau_{\rm R}
\sim -0.5$ are, however, stable to convection. Hence Fig.~\ref{fg:f12} suggests
that these layers may take up to 100 seconds to reach radiative
equilibrium. The radiation field becomes weaker in cooler white dwarfs, and
the decreasing line opacity creates upper layers that do not interact much
with radiation. This contributes in creating the long observed radiative
timescales. A similar phenomenon was observed for metal-poor dwarfs
\citep{asplund01}.

To make sure that our simulations have relaxed in the upper layers, we
performed non-grey 2D simulations for 50 seconds for the three models between
7000 and 9000~K, and for 90 seconds for the model at 6000~K. The upper layers
never actually reach a radiative equilibrium like in the 1D models. Instead
the convective overshoot causes the entropy gradient in the upper layers to
relax to a near-adiabatic structure, as seen in Fig.~\ref{fg:f8}. We used the
final 2D structures as initial conditions\footnote{We copied all quantities of
  the 2D snapshots in the third dimension, except for the velocities which
  were taken from a previously computed 3D snapshot.} for the much more time
consuming non-grey 3D simulations.

All non-grey 3D simulations were run for 10 seconds, which ensures that we
typically cover about 100 advective timescales in the photosphere. We have
verified that all simulations are relaxed in the last five seconds of
computation and that they show no systematic and non-oscillatory change of
their properties in all layers. We note that the four cooler simulations might
still have an imprint of the 2D initial conditions in the upper layers because
of the large radiative timescales.

The numerical time step is limited by the global minimum of the radiative and
Courant-Friedrichs-Lewy timescales ($t_{\rm CFL} = \Delta x /[c_s + v_c]$,
where $c_s$ the adiabatic sound speed). The latter corresponds the travel time
of the shortest wave across a grid cell of dimension $\Delta x$. Typically,
our time steps are of the order $10^{-4}-10^{-5}$ second, and therefore the
total number of steps is $t_{\rm sim}/\Delta t \sim 10^{5}-10^{6}$.

\subsection{Characteristic length scales}

We have previously discussed the vertical extent of our simulations which is
naturally derived from our requirement to have the full atmospheres. However,
the horizontal dimensions of the simulations are not well defined a priori.
The 3D simulations have horizontal sizes, while such thing does not exist in 1D
plane-parallel models. From physical considerations in the formation,
transport and dissipation of convective cells, one can assume that the
  vertical size of the cells will be of the order of the local pressure scale
  height. However, calculations of stellar RHD models have shown that granular
  patterns have larger horizontal dimensions of the order of 10 times the
  local value of $H_{\rm p}$ \citep{freytag97}.

\begin{figure}[!h]
\begin{center}
\includegraphics[bb=18 165 672 652,width=4in]{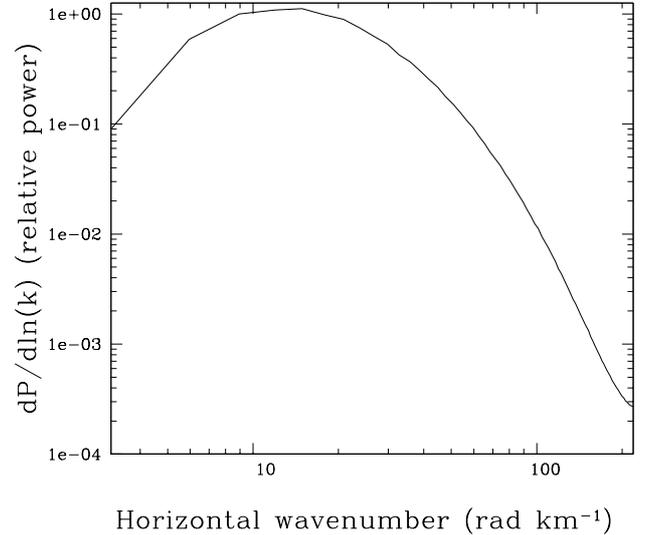}
\caption{Mean power spectrum as a function of the horizontal wavenumber
  (2$\pi$/$\lambda$) averaged over 250 snapshots of the (bolometric) intensity
  map of the 10,000~K 3D simulation.
\label{fg:f13}}
\end{center}
\end{figure}

In Fig.~\ref{fg:f13}, we present the power spectrum as a function of the
horizontal wavenumber, for the emergent intensity in 250 different snapshots
of the 10,000~K simulation (Fig.~\ref{fg:f4} gives one example of a
snapshot). We display power per logarithmic wavenumber interval for a
  more direct identification of the power carrying scales
  \citep{ludwig02}. We remind the reader that the smallest wavenumber
on the left in Fig.~\ref{fg:f13} represents a sinusoidal pattern with one hot
crest and one cool crest (compared to the mean intensity). Ideally, our
simulations should include at least of the order of $3\times3$ hot cells to
have a good representation of the atmosphere, and have a sufficient resolution
for each cell. We only qualitatively ensured that it was the case since
  we can derive the power spectra only after the simulations. The
  characteristic granulation is typically well resolved although because of
  the limitation of the Fourier analysis to identify granules and possible
  effects from the numerical parameters, the characteristic dimensions should
  be taken as estimates only. We note that we can not derive a single power
law to describe the formation of smaller cells or sub-cells with a high
wavenumber, much like in main-sequence simulations \citep{ludwig02}.

\begin{figure}[!h]
\begin{center}
\includegraphics[bb=18 165 672 652,width=4in]{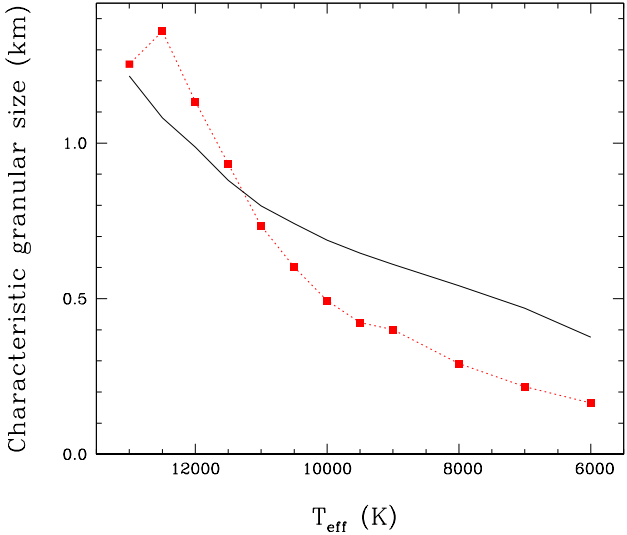}
\caption{Characteristic granular size (maximum of the mean power spectrum) as
  a function of $T_{\rm eff}$ for our sequence of 3D simulations. The points are
  connected for clarity. The solid line is eight times the pressure scale
  height at $\tau_{\rm R}$ = 2/3.
\label{fg:f14}}
\end{center}
\end{figure}

Fig.~\ref{fg:f14} presents the characteristic size of the convective cells in
our sequence of simulations, derived as the wavelength of the peak of the
power spectra which we found were well fitted by a parabola. It is clearly
seen that the cell dimensions are not well scaled by the pressure scale height
in the photosphere, also shown in the figure. The characteristic size of the
granulation patterns increases much more rapidly with $T_{\rm eff}$ than what
we would expect from simple thermodynamics considerations. The convective
velocities and the associated Mach number are significantly increasing with
$T_{\rm eff}$, and we suggest that it has an effect on the size of the
convective cells by allowing for faster moving and thinner
downdrafts. Furthermore, the timescales analysis of Sect.~3.3 suggests that
the variation of the P\'eclet number is part of the explanation. The regime of
regular bright and dark cells at cool temperatures (see, e.g.,
Fig.~\ref{fg:f3}), where the advection dominates, change to the regime of
large bright cells and narrow dark lanes at hotter temperatures
(Fig.~\ref{fg:f5}) when the radiation dominates.

\begin{figure}[!h]
\begin{center}
\includegraphics[bb=18 165 672 652,width=4in]{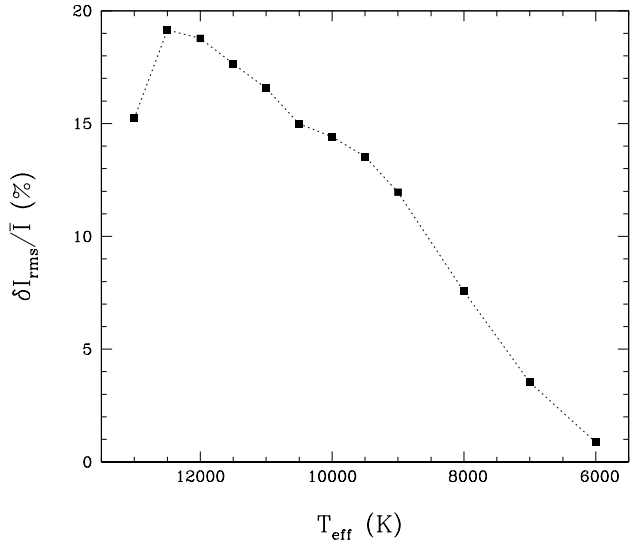}
\caption{RMS intensity contrast divided by the mean intensity as a function of
  $T_{\rm eff}$ for our sequence of 3D simulations. The points are connected for
  clarity.
\label{fg:f15}}
\end{center}
\end{figure}

Also of interest is the intensity contrast of the granulation patterns with
respect to the mean intensity. This quantity can be derived from the intensity
snapshots such as the ones shown in Figs. \ref{fg:f3}-\ref{fg:f5}. The RMS
intensity contrast, a 3D effect that can not be easily predicted from existing
1D structures, is presented in Fig.~\ref{fg:f15}. It is seen that the contrast
varies from 1 to 20$\%$ in our sequence. In all but the three coolest models,
the intensity contrast is in the 10-20$\%$ range. Very similar values of the
intensity contrast are found for F, G and K dwarfs \citep[see Fig.~15
  of][]{freytag12}. We find that the granulation is visually similar in the
Sun and a $\sim$11,000~K white dwarf. In cooler white dwarfs, the granulation
appears less structured in comparison to K main-sequence stars (see
Fig.~\ref{fg:f3}). We note that cool white dwarfs have a wide entropy
  minimum, which implies that the upper boundary of the convective zone is not
  well defined and that granulation patterns may form in a more extended
  optical depth range than dwarfs. This is a possible explanation for the
  fuzziness of granulation in cool white dwarfs. A careful study of the
granulation patterns as a function of $T_{\rm eff}$, $\log g$ and metallicity
for stars and white dwarfs would be useful to understand these structures.

We mention that at 13,000~K, the intensity contrast is still high, while the
convective flux is fairly low according to Fig.~\ref{fg:f9}. It appears that
because the total energy flux is large in the atmosphere, large temperature
differences are necessary to transport even a small amount of convective flux.
However, the differences between 1D and $\langle {\rm 3D}\rangle$ models are
small according to Fig.~\ref{fg:f7}. This may not be entirely surprising,
since the weak convective flux is unlikely to have a global impact on the
largely radiative structures. At the opposite cool end of our sequence, the
6000~K simulation features an intensity contrast of about 1$\%$. It suggests again
that convection is very efficient to transport the small total flux in this
regime with only very small temperature variations. In between these two
extremes, the increasing total flux contributes to enhance the
contrast. Furthermore, as $T_{\rm eff}$ is increasing, the Rosseland opacity
is also increasing in the photospheres. This implies lower characteristic
densities in the photospheres and higher convective velocities. This effect
also contributes in enhancing the fluctuations.

Finally, we recall here that despite the significant variations in the
spatially resolved intensity or flux, results of Paper I have shown that the
mean wavelength-dependent flux is almost exactly equal to the same quantity
computed from mean $\langle {\rm 3D}\rangle$ structures. Hence the 3D
fluctuations impact only indirectly the predicted spectra of white dwarfs
through global modifications of the structures. In the following, the 3D
simulations are therefore represented by their mean structures.

\section{Astrophysical applications and discussion}

\subsection{Model spectra}

\begin{figure*}[!ht]
\begin{center}
\includegraphics[bb=88 118 522 712, width=3.5in,angle=270]{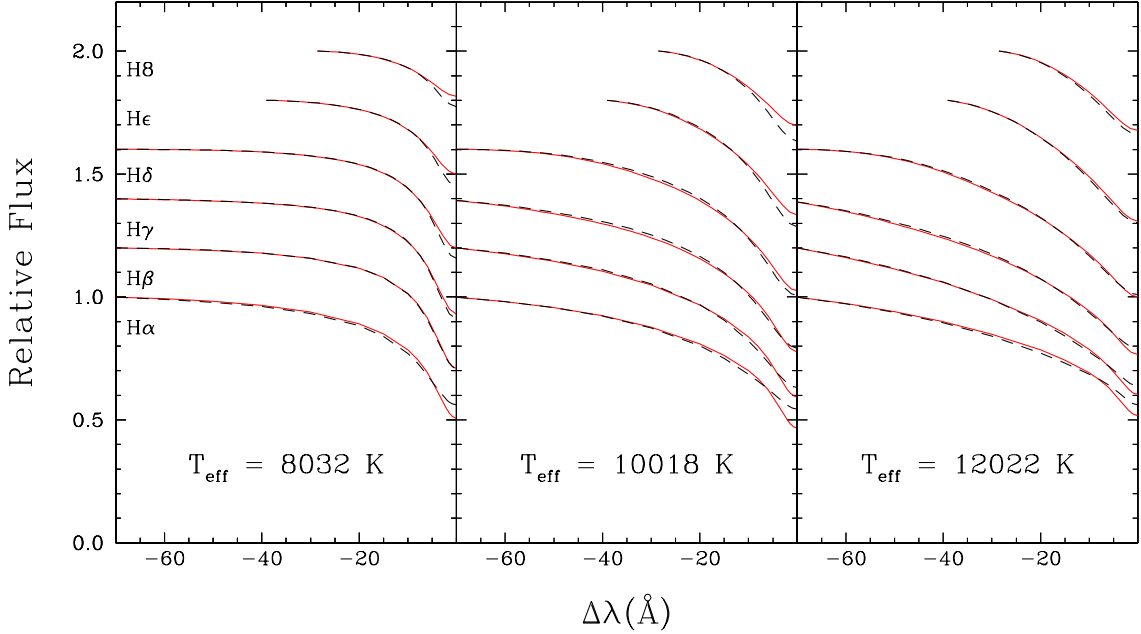}
\caption{Comparison of the blue wing of six Balmer line profiles (H$\alpha$ to
  H8) calculated from $\langle {\rm 3D}\rangle$ structures (red, solid) and
  standard 1D structures (black, dashed) for three models from our sequence
  identified in Table~2. $T_{\rm eff}$ values are shown on the different
  panels. All line profiles were normalized to a unity continuum at a fixed
  distance from the line center. All spectra were convolved with a Gaussian
  profile with a resolution of 6 $\AA$ to represent typical observations.
\label{fg:f16}}
\end{center}
\end{figure*}

Our sequence of 3D model atmospheres can be used as input for spectral
synthesis as done in Paper I. We have shown in Sect.~2 that our 3D simulations are
based on the same input microphysics as the standard 1D models, hence we can
use the 1D model atmosphere code of \citet{TB11} to compute $\langle {\rm
  3D}\rangle$ spectra, from $\langle {\rm 3D}\rangle$ structures, that can be
compared directly to observations. However, since our sequence of 3D calculations is
limited to one $\log g$ value, we can not fit actual observations and we will
still rely on a $\langle {\rm 3D}\rangle$$-$1D differential approach to look
at the 3D effects. We compare in Fig.~\ref{fg:f16} the $\langle {\rm
  3D}\rangle$ and 1D spectra for three characteristic simulations
from our sequence of model atmospheres identified in Table~2. For clarity,
only the blue wings of the lines are shown since the comparison is similar for
the red wings. We must conclude, like in Paper I, that the differences are
fairly subtle between the predicted $\langle {\rm 3D}\rangle$ and 1D spectra,
and that we must be careful about drawing conclusions about the atmospheric
parameter corrections.

We find that the largest differences between $\langle {\rm 3D}\rangle$ and 1D
spectra are in the middle of our sequence, at $T_{\rm eff} \sim 10,000$~K,
which is also the region where the structures are the most different (see
Fig.~\ref{fg:f7}) and where the high-$\log g$ problem is the most significant
(see Fig. 1 of Paper I). One surprising result is that the higher lines of the
series are significantly impacted by 3D effects. Since the higher series
members are formed in a narrow region of the photosphere, one would expect
that 3D effects decrease for these lines. The reason for this behaviour is
that the strength of the higher lines is sensitive to the non-ideal effects
\citep{hm88,TB09} in comparison to the lower lines. The non-ideal effects are
in turn sensitive to the density in the atmospheres, and $\langle {\rm
  3D}\rangle$ structures have systematically cooler temperatures and higher
densities in the upper layers due to the convective overshoot. These higher
densities enhance the non-ideal effects and reduce the line strengths in
comparison to the 1D predictions.

The cores of H$\alpha$ and to a lesser degree H$\beta$ are significantly
deeper in $\langle {\rm 3D}\rangle$ spectra. These cores are formed very high
in the atmospheres, where the $\langle {\rm 3D}\rangle$ structures deviate
significantly from their 1D counterparts due to the overshoot cooling. Since
spectroscopic analyses of white dwarfs with 1D models provide good fits to the
line centers, the predictions from the $\langle {\rm 3D}\rangle$ spectra
should be regarded with caution. We believe that the lower temperatures
predicted by our 3D simulations are real, although we can not rule out that
shortcomings in radiative transfer or missing physics (e.g. magnetic fields or
shock formation) might have an impact on this issue. We have verified with the
TLUSTY code \citep{TLUSTY} that the 1D NLTE effects are restricted to the very
center of the lower lines ($\Delta \lambda < 1 \AA$) in this regime of $T_{\rm
  eff}$ and that they contribute to further enhance the absorption. Therefore,
it seems unlikely that NLTE effects can help in explaining the differences
between $\langle {\rm 3D}\rangle$ and 1D line cores. As a conservative
measure, we remove the line centers ($\vert \Delta \lambda \vert < 1.5~\AA$) in our
derivation of the 3D atmospheric parameter corrections.

\subsection{Application to the high-log g problem}

\begin{figure}[!h]
\begin{center}
\includegraphics[bb=18 165 672 652,width=4in]{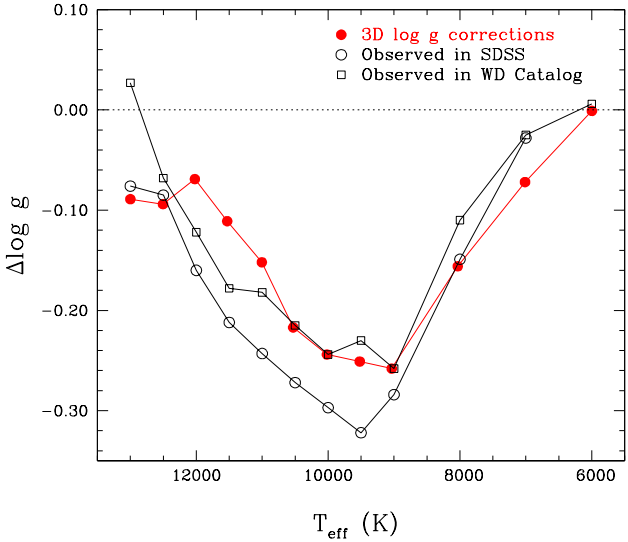}
\caption{3D $\log g$ corrections (red filled circles) as a function
  of $T_{\rm eff}$. In comparison, we show the shifts in surface gravity
  required to obtain a stable $\log g$ distribution as a function of $T_{\rm
    eff}$ using the samples of the SDSS (open circles) and Gianninas et
  al.~(2011; open squares). The points are connected for clarity.
\label{fg:f17}}
\end{center}
\end{figure}

We derive here 3D atmospheric parameter corrections that are defined as the
differences in $T_{\rm eff}$ and $\log g$ when we fit the normalized line
profiles of the $\langle {\rm 3D}\rangle$ spectra with our grid of standard 1D
models. More specifically, the corrections are defined as 1D$-$$\langle {\rm
  3D}\rangle$, since we are interested in how much the atmospheric parameters
of real stars would change when using $\langle {\rm 3D}\rangle$ model
spectra. In Fig.~\ref{fg:f17} and in Table~2, we present the 3D $\log g$
corrections. As in Paper I, we find that the corrections are negative for all
spectra (i.e., the 3D simulations predict lower surface gravities), which is in the
appropriate direction to correct the high-$\log g$ problem. We note that the
3D $\log g$ corrections are smaller in absolute value than those found in
Paper I in the range of $T_{\rm eff}$ in common between both studies, i.e. the
four hottest simulations. This is a direct consequence of the fact that the 1D LHD
models used in Paper I had optically thick convective cells everywhere in the
atmosphere and were inconsistent with standard 1D models (see Sect.~2.2).

Also given in Fig.~\ref{fg:f17} are the observed shifts in $\log g$ derived
from the spectroscopic analyses of the DA white dwarfs in the Sloan Digital
Sky Survey \citep{TB11} and White Dwarf Catalog \citep{gianninas11}. These shifts
correspond to the corrections required, in a bin of 1000~K around the simulation
temperature, to match the mean mass value obtained from hot DA stars. We can
see that the 3D corrections describe well the shape of the high-$\log g$
problem, with maximum corrections at $T_{\rm eff} \sim 10,000$~K where the
problem is the largest. Also, the 3D corrections have about the right
amplitude everywhere to solve the high-$\log g$ problem. Knowledge of the 3D
$\log g$ corrections at $\log g$ = 7.5 and 8.5, and ultimately the
spectroscopic fit of observations will be necessary to further constrain the
$\log g$ values predicted by the 3D simulations. However, our results confirm the
conclusion of Paper I that 1D MLT convection is the main reason for the
high-$\log g$ problem, and that 3D model atmospheres provide a more stable
surface gravity distribution as a function of the effective temperature.

We find that the 3D $T_{\rm eff}$ corrections are mild, with an average of
$-$230~K for the simulations in our sequence, hence the 3D effects are mostly $\log
g$ effects. We believe that an actual comparison with observations will be
necessary to interpret further the $T_{\rm eff}$ shifts.

\subsection{Sensitivity to numerics}

In order to understand the precision of our 3D simulations and the uncertainty
of our $\log g$ corrections, we computed a series of 9 simulations with the
same input parameters, except for one modification as given Table~3. All
simulations were computed at $T_{\rm eff} \sim 10,000$~K and $\log g = 8$ for
5 seconds (at least one order of magnitude larger than both $t_{\rm rad}$ and
$t_{\rm adv}$ in the photosphere), with the converged model from our regular
sequence as the starting model. We derived the mean temperature and pressure
structures from 6 snapshots, and computed spectra using the same approach as
we did for our regular sequence. In Table~3, we present how the 3D $\log g$
corrections are changed by our modifications.

 \begin{table}[!h]
 \caption{Sequence of 3D simulations with alternative parameters}
 \label{tab3}
 \begin{center}
 \begin{tabular}{lr}
\hline 
\hline 
Description & $\log g$ shift \\
\hline
A) Surface area $A_{xy}$ = 1.56 $A_{xy,0}$ & 0.00 \\
B) Vertical grid points $n_z$ = 2/3 $n_{z,0}$  & 0.00 \\ 
C) $\log \tau_{max}$ = 0.93 $\log \tau_{max, 0}$ & -0.01 \\
D) Top layer inflow $T = 1.1 \times T_{0}$ & 0.00 \\
E) Van Leer slope limiter & -0.00 \\
F) Piecewise Parabolic reconstruction & -0.02 \\
G) Artificial viscosity & 0.00 \\
\hline
H) 2D calculation & 0.06 \\
\hline
I) Original opacity table (Ludwig et al. 1994) & 0.03 \\
\hline

\end{tabular} 
\end{center} 
\tablefoot{All models with $T_{\rm eff} \sim 10,000$~K and $\log g = 8$. The
  subscript $0$ refers to the standard model parameters.}.
\end{table}

The first category of modifications (simulations A to G) can be classified as
tolerable, i.e. from the physical constraints derived in this paper, these
could be valid alternative input parameters for models in the regular
sequence. We have A) increased the horizontal geometrical dimensions $x$ and
$y$ by 25\% (hence lowered the horizontal resolution), B) decreased the
vertical resolution by using 100 grid points instead of 150, C) cut a section
of atmosphere by changing the maximum value of $\log \tau_{\rm R}$ from 3.0 to 2.8,
D) changed the top boundary condition by increasing by 10\% the temperature of
the incoming flow, E) used the Van Leer slope limiter, as in Paper I, instead
of the less dissipative 2nd order method, F) used the more aggressive
Piecewise Parabolic reconstruction \citep{colella84} and G) added artificial
viscosity as in Paper I. According to Table~3, all of these modifications have
no significant effect on the predicted gravities. This is a very important
result, which confirms that the unresolved turbulent energy dissipation is not
an issue for the predicted Balmer lines.

The simulations H and I include modifications that presumably result in a
significantly less accurate account of the physical processes in comparison to
the models in our regular sequence. We computed a 2D simulation by removing
one of the horizontal direction. We find that the $\log g$ correction is
changed by +0.06 dex, which is a moderate but significant effect. This result
is meaningful since 2D models have been used in this work to provide relaxed
upper layers for cool simulations (see Sect.~3.3). In fact, since 2D
simulations are an order of magnitude faster to compute than 3D models, they
are an interesting resource for applications where the geometrical dimensions
or calculation times are larger than those in our sequence of 3D
simulations. While these 2D calculations may not be as accurate as the 3D
setup to predict the Balmer line spectra, they still appear to provide a very
good physical description of the atmospheres. We note that early works on RHD
models of white dwarfs were done with 2D simulations. Our result shows that
their 2D approximation was adequate to derive white dwarf properties.

We close this section with the simulation I where we relied on the opacity
table first computed by \citet{ludwig94} and also used in Paper I. We find
that the $\log g$ correction is only changed by +0.03 dex. Therefore, we
conclude that the opacity binning procedure was already very well tuned in
previous works on RHD models of white dwarfs.

\section{Conclusion}

We computed a sequence of twelve 3D model atmospheres with the CO$^5$BOLD
radiation-hydrodynamics code. These simulations were made in the range $13,000
>$ $T_{\rm eff}$ (K) $> 6000$ and at $\log g = 8$ for pure-hydrogen
atmospheres. We relied on EOS and opacity tables that have the same
microphysics as in standard 1D models of white dwarfs, and we have verified
that our models are expected to have the same precision as prevalent 1D
structures. As a consequence, our sequence of 3D simulations is now expected
to predict realistic absolute properties and can ultimately be compared with
observations. We have demonstrated that the 3D simulations depend only weakly
on numerical parameters by running a set of alternative simulations with
modified parameters. This is unlike 1D models for which the free parameters in
the mixing-length theory have a significant impact on the predicted
structures.

We derived $\langle {\rm 3D}\rangle$ spectra of the Balmer lines that we
compared with those predicted from standard 1D white dwarf models. We found
that the 3D $\log g$ corrections are the largest at around $T_{\rm eff} \sim
10,000$~K and that they have the right amplitude as a function of $T_{\rm
  eff}$ to draw the conclusion that the source of the long-standing high-$\log
g$ problem is the inability of the mixing-length theory to properly account
for the convective energy transport. We will follow this study with the
calculation of sequences of simulations at $\log g$ = 7.0, 7.5, 8.5 and
9.0. It will then be possible to apply the 3D model atmospheres in the
spectroscopic analysis of DA white dwarf samples.

\begin{acknowledgements}
P.-E. T. is supported by the Alexander von Humboldt Foundation. 3D model
calculations have been performed on CALYS, a mini-cluster of 320 nodes built
at Universit\'e de Montr\'eal with the financial help of the Fondation
Canadienne pour l'Innovation. We thank Prof. G. Fontaine for making CALYS
available to us. We are also most grateful to Dr. P. Brassard for technical
help. This work was supported by Sonderforschungsbereich SFB 881 "The Milky
Way System" (Subproject A4) of the German Research Foundation
(DFG). B.F.\ acknowledges financial support from the {\sl Agence Nationale de
  la Recherche} (ANR), and the {\sl ``Programme Nationale de Physique
  Stellaire''} (PNPS) of CNRS/INSU, France.
\end{acknowledgements}

\bibliographystyle{aa} 

\begin{thebibliography}{}

\bibitem[Allard \& Kielkopf(1982)]{allard82} Allard, N.~F., \& Kielkopf, J.~F.\ 1982, Reviews of Modern Physics, 54, 1103 

\bibitem[Allard et al.(2004)]{allard04} Allard, N.~F., Kielkopf, J.~F., \& Loeillet, B.\ 2004, \aap, 424, 347 

\bibitem[Asplund \& Garcia Perez(2001)]{asplund01} Asplund, M., \& Garcia Perez, A. E.\ 2001, \aap, 372, 601

\bibitem[Asplund et 
al.(2009)]{asplund09} Asplund, M., Grevesse, N., Sauval, A.~J., \& Scott, P.\ 2009, \araa, 47, 481 

\bibitem[Bergeron et al.(1990)]{bergeron90} Bergeron, P., Wesemael, F., Fontaine, G., \& Liebert, J.\ 1990, \apjl, 351, L21 

\bibitem[Bergeron et al.(1992a)]{bergeron92a} Bergeron, P., Saffer, R.~A., \& Liebert, J.\ 1992a, \apj, 394, 228

\bibitem[Bergeron et al.(1992b)]{bergeron92b} Bergeron, P., Wesemael, F., \& Fontaine, G.\ 1992b, \apj, 387, 288 

\bibitem[Bohlin(2000)]{bohlin00} Bohlin, R.~C.\ 2000, \aj, 120, 
437 

\bibitem[B\"ohm-Vitense(1958)]{MLT} B\"ohm-Vitense, E.\ 1958, ZAp, 46, 108

\bibitem[Brassard \& Fontaine(1997)]{brassard97} Brassard, P., \& Fontaine,
  G.\ 1997, in Proc. 10th European Workshop on White Dwarfs, eds. J. Isern,
  M. Hernanz \& E. Gracia-Berro (Dordrecht: Kluwer), 214, 451

\bibitem[Caffau \& Ludwig(2007)]{caffau07} Caffau, E., \& Ludwig, H.-G.\ 2007, \aap, 467, L11

\bibitem[Caffau et al.(2011)]{caffau11} Caffau, E., Ludwig, 
H.-G., Steffen, M., Freytag, B., \& Bonifacio, P.\ 2011, \solphys, 268, 255 

\bibitem[Colella(1990)]{colella90} Colella, P.\ 1990, Journal of 
Computational Physics, 87, 171 

\bibitem[Colella(1984)]{colella84} Colella, P., \& Woodward, P.~R.\ 1984, \jcp, 54, 174

\bibitem[Eisenstein et al.(2006)]{SDSS} Eisenstein, D.~J., Liebert, J., Harris, H.~C., et al.\ 2006, \apjs, 167, 40 

\bibitem[Finley et al.(1997)]{finley97} Finley, D.~S., Koester, D., \& Basri, G.\ 1997, \apj, 488, 375 

\bibitem[Fontaine \& Brassard(2008)]{fontaine08} Fontaine, G., \& Brassard, P.\ 2008, \pasp, 120, 1043 

\bibitem[Freytag et al.(1996)]{freytag96} Freytag, B., Ludwig, H.-G., \& Steffen, M.\ 1996, \aap, 313, 497

\bibitem[Freytag et al.(1997)]{freytag97} Freytag, B., Holweger, H., Steffen,
  M., \& Ludwig, H.-G.\ 1997, in Science with the VLT Interferometer,
  ed. F. Paresce (New York: Springer), 316

\bibitem[Freytag et al.(2012)]{freytag12} Freytag, B., Steffen, 
M., Ludwig, H.-G., et al.\ 2012, Journal of Computational Physics, 231, 919 

\bibitem[Gautschy et 
al.(1996)]{gautschy96} Gautschy, A., Ludwig, H.-G., \& Freytag, B.\ 1996, \aap, 311, 493 

\bibitem[Gianninas et al.(2011)]{gianninas11} Gianninas, A., 
Bergeron, P., \& Ruiz, M.~T.\ 2011, \apj, 743, 138 

\bibitem[Girven et al.(2011)]{girven11} Girven, J., 
G{\"a}nsicke, B.~T., Steeghs, D., \& Koester, D.\ 2011, \mnras, 417, 1210 

\bibitem[Hubeny \& Lanz(1995)]{TLUSTY} Hubeny, I., \& Lanz, T.\ 1995, \apj, 439, 875 

\bibitem[Hummer 
\& Mihalas(1988)]{hm88} Hummer, D.~G., \& Mihalas, D.\ 1988, \apj, 331, 794 

\bibitem[Koester et 
al.(1994)]{koester94} Koester, D., Allard, N.~F., \& Vauclair, G.\ 1994, \aap, 291, L9 

\bibitem[Koester et al.(2009a)]{koester09a} Koester, D., Kepler, S. O., Kleinman, S. J., \& Nitta, A.\ 2009a, J. Phys.: Conf. Ser., 172, 012006

\bibitem[Koester et al.(2009b)]{koester09b} Koester, D., Voss, B., Napiwotzki, R., et al.\ 2009b, \aap, 505, 441 

\bibitem[Kowalski \& Saumon(2006)]{kowalski06} Kowalski, P.~M., \& Saumon, D.\ 2006, \apjl, 651, L137 

\bibitem[Liebert et al.(2005)]{LBH05} Liebert, J., Bergeron, P., \& Holberg, J.~B.\ 2005, \apjs, 156, 47 

\bibitem[Ludwig et al.(1994)]{ludwig94} Ludwig, H.-G., Jordan, S., \& Steffen, M.\ 1994, \aap, 284, 105 

\bibitem[Ludwig et al.(1999)]{ludwig99} Ludwig, H.-G., Freytag, B., \& Steffen, M.\ 1999, \aap, 346, 111 

\bibitem[Ludwig et al.(2002)]{ludwig02} Ludwig, H.-G., Allard, F., \& Hauschildt, P.~H.\ 2002, \aap, 395, 99 

\bibitem[Ludwig \& Ku{\v c}inskas(2012)]{ludwig12} Ludwig, H.-G., \& Ku{\v c}inskas, A.\ 2012, \aap, 547, A118 

\bibitem[Mihalas(1978)]{mihalas78} Mihalas, D.\ 1978, San Francisco, W.~H.~Freeman and Co., 1978.~650 p

\bibitem[Nordlund(1982)]{nordlund82} Nordlund, {\AA}.\ 1982, \aap, 107, 1 

\bibitem[Steffen et al.(1995)]{steffen95} Steffen, M., Ludwig, H.-G. \& Freytag, B.\ 1995, \aap, 300, 473

\bibitem[Tassoul et al.(1990)]{tassoul90} Tassoul, M., Fontaine, 
G., \& Winget, D.~E.\ 1990, \apjs, 72, 335 

\bibitem[Tremblay \& Bergeron(2009)]{TB09} Tremblay, P.-E., \& Bergeron, P.\ 2009, \apj, 696, 1755

\bibitem[Tremblay et al.(2010)]{TB10} Tremblay, P.-E., Bergeron, P., Kalirai, J.~S. \& Gianninas, A.\ 2010, \apj, 712, 1345

\bibitem[Tremblay et al.(2011a)]{TB11} Tremblay, P.-E., Bergeron, P. \& Gianninas, A.\ 2011a, \apj, 730, 128

\bibitem[Tremblay et 
al.(2011b)]{paper1} Tremblay, P.-E., Ludwig, H.-G., Steffen, M., Bergeron, P., \& Freytag, B.\ 2011b, \aap, 531, L19 (Paper I)

\bibitem[Vitense(1953)]{vitense53} Vitense, E.\ 1953, \zap, 32, 
135 

\bibitem[V{\"o}gler et 
al.(2004)]{voegler04} V{\"o}gler, A., Bruls, J.~H.~M.~J., \& Sch{\"u}ssler, M.\ 2004, \aap, 421, 741 

\bibitem[Weidemann \& Koester(1980)]{weidemann80} Weidemann, V. \& Koester, D.\ 1980, \aap, 85, 208

\bibitem[van Grootel et 
al.(2012)]{grootel12} van Grootel, V., Dupret, M.-A., Fontaine, G., et al.\ 2012, \aap, 539, A87 

\end{thebibliography}

\end{document}